# Timeless Relativistic Approach to Classical Mechanics

Samuel H. Talbert*

**Abstract**

We reformulate Classical Mechanics as a timeless relativistic theory. Readers are introduced to a new class of reference systems, the binate frames, where physical events are identified with four position-coordinates -- no clocks are used. The binate frames are inertial and adaptable to relative motions. Analyses that use binate frames are valid at all energy levels. When desirable to do so, the results are easily expressed as in special relativity, in terms of space and time coordinates of inertial observers. Given the importance of mass-and-spring systems to theoretical physics, we analyzed such a system. Published special relativistic solutions predict anharmonic oscillations, but in the inertial frame where the mass-and-spring system (as a whole) is at rest, our solution is harmonic with energy-dependent angular frequency.

* **Electronic address: talberts@copper.net**

> "To understand a subject, one must tear it apart and reconstruct it in a form intellectually satisfying to oneself, and that (in view of the differences between individual minds) is likely to be different from the original form."
>
> John Layton Synge [1]

# I. INTRODUCTION

To the Minkowskian view that the only independent realty is a kind of union of space and time, we oppose a Cartesian view where three-dimensional spaces exist only as the indefinite extension of material bodies, and where the relative space of each body moves in the relative spaces of all other bodies. This Cartesian view is the foundation on which we built a timeless relativistic approach to classical mechanics. Since the new approach uses the same mathematical formalism as classical mechanics and special relativity, the presentation spotlight is here on the physical aspects of the new approach. For all intent and purposes, the article is a tutorial on how the use of a new class of inertial reference systems, the *binate frames*, endows classical mechanics with the full predictive power of special relativity.

Briefly, a binate frame is formed with two distinct inertial frames of the kind used in classical mechanics. Such frames carry Cartesian coordinates but no clocks. In a binate frame, a physical event is identified with the four position-coordinates of places from the extended spaces of material bodies -- there is no time-coordinate. Canonical equations of motion are derived from a system Hamiltonian and their solution yields the paths of the system particles through phase space. If and when desirable to do so, these paths could be presented in Newtonian or Minkowskian spacetime by stipulating a time-coordinate for each inertial frame of the pair that forms the binate frame. The differences between the two mentioned spacetimes arise from how their respective time-coordinate is defined. For the Newtonian spacetime, the path parameter provides a common time-coordinate for the two inertial frames. For the Minkowskian, two distinct time-coordinates are employed, one for each inertial frame. The beauty of the binate-frame approach is that the canonical equations of motion are derived and solved without a commitment to a time-coordinate, a commitment in fact unnecessary even for the experimental verification of the solution.

The availability of binate frames in no way denies the practical need to coordinate human activities with synchronized clocks. But why use a time-coordinate to explain observed phenomena and to predict the outcome of experiments? It is very tempting to immediately reply that a time-coordinate is obviously needed to analyze the phenomena and experiments that so far could only be explained with the special theory of relativity. But we show here that binate frame analyses predict the null-effect of Michelson-Morley experiment (Sec. IVB), the decay of cosmic muons in flight (Sec. VIC), the Wigner angle (Sec. VII), and the Bertozzi relativistic energy curve obtained in experiments with high-speed electrons (Sec. VIIIC). These should suffice to convince the reader that the binate frame approach is a valid timeless alternative to special relativity. We present below the considerations that led us to the notion of a binate frame, and follow with an overview of the article.

In classical mechanics, a frame of reference is an origin together with a set of right-handed Cartesian coordinate axes. The classical 'time' is just an independent parameter.

While it is generally understood that such a frame defines a *standard of rest*, it does not seem to have been noticed that a pair of such classical inertial frames defines not only a standard of *direction* (the line of relative motion is a baseline for measuring angles) but also a *standard of motion*. This is easily explained: There is ample evidence that when two inertial frames are in relative motion, the frequency of a light signal detected on one of them is not the same as the frequency emitted on the other. This optical Doppler effect is evidently a measure of the relative motion between two inertial frames. In this sense, a pair of classical inertial frames provides a standard for any other relative motion. Could the pair be used as a new kind of reference frame? And if this were so, what is the group of transformations between the binate frames? Is it isomorphic with Lorentz group? And if the two groups were indeed isomorphic, what is the mapping between them? Finally, if our questions were to have positive answers, would modern theoretical physics benefit from using this new type of reference system formed with two classical inertial frames in relative motion? The article offers partial answers to the posed questions, for a critical review by its readers.

Following this introduction, we first present in Sec. II some simplifying assumptions that our approach has in common with classical and relativistic kinematics. Experiments with light signals enable the specification of a $k$-ratio of two distances that we use here to measure the local rate at which coinciding places from two bodies in relative motion separate from each other. We show that this ratio equals numerically the longitudinal Doppler factor for light: $k=\sqrt{(1-\beta)/(1+\beta)}$. The relation between the here defined $k$ and the relative speed $\beta$ is a bridge that we cross often during the presentation, but only for the purpose of comparing our results with those of special relativity.

The relative motion between two classical inertial frames identifies on each frame a preferred direction. We show in Sec. III that a pair of classical inertial frames could then be used as a reference system, a *binate frame*. This is characterized by a constant $k$-ratio and a preferred direction. The equation of a wave front from an emitted light, which we derive in Sec. IV, is the binate frame counterpart of the light cone. It involves the $k$-ratio and, on each of the two inertial frames that form the specified binate frame, two Cartesian vectors: (i) a unit vector in the preferred direction and (ii) a displacement vector from the place of emission to that of detection. Our derivation rests on the evidence that material bodies (with negligible gravitational fields) do not affect the propagation of a light signal when they are not obstructing its path.

We show in Sec. V how two selected events identify canonical Cartesian coordinates on each of the two inertial frames used to form a binate frame. In terms of these canonical coordinates, the matrix $\eta_{\mu\nu}(k)$ of our light propagation quadric, $\eta_{\mu\nu}dx^{\mu}dx^{\nu}=0$, emerges symmetric and with the expected $(+,-,-,-)$ signature, but not diagonal. We refrain from invoking coordinate changes that brings our matrix to its diagonal form, and thus enables a spacetime interpretation, since this form is not unique (though its signature is).

Evidently, there is no notion of 'rest' in a binate frame: a massive particle, not unlike a photon, is always in motion relative to at least one of the two inertial frames that form the binate frame. Based on the ample evidence that a massive particle could not see twice an emitted light (unless of course the light is reflected back toward the particle), we show in Sec. VI that the separation interval, $ds^2 = \eta_{\mu\nu} dx^\mu dx^\nu$ , between near events attended by the particle is always positive. We prove its invariance when changing the canonical coordinates (Sec. VIA), and when changing the binate frame (Sec. VIB).

The kinematical part of our presentation finishes in Sec. VII with an analysis that explains the Wigner angle effect in a three-body situation.

The binate frames are devoid of a kinematic notion of simultaneity of distant events. But in a multi-particle system, a dynamic notion of simultaneity emerges from the path parameter of the system evolution through phase space. The parameter is the independent variable of the canonical equations of motion, which are derived here (Sec. VIII) from Hamiltonians expressed in terms of canonical position-coordinates and their conjugate momentum-coordinates. To be physically meaningful, such Hamiltonians should have the same form and value in all binate frames, regardless of their canonical parameterization. This then ensures that also the path parameter is invariant, and thus could provide the multi-particle system with a notion of distant simultaneity.

In practice, it is possible to simplify the calculations by using a binate-frame adapted to some relative motion between the particles of the considered system. We illustrate this with an analysis in Sec. IX of the dynamics of a constrained two-body system. Each body moves on a track that makes a right angle with the track on which the other body moves. The two bodies cannot both move uniformly since a massless rigid rod connects them.

When desirable to do so, the results of a binate frame approach could be rewritten in terms of the spacetime coordinates of inertial observers. We do the rewriting often during the presentation, for comparison purposes. Occasionally, we get a solution that differs from all the published special relativistic solutions. We included an example in Sec. XA: In view of the basic nature of the relativistic oscillator in modern physics, we analysed a mass-and-spring system. Our analysis, which uses an adapted binate-frame, predicts that the mass moves harmonically in the inertial frame where the mass-and-spring system, as a whole, is at rest. The angular frequency of this motion is energy-dependent. In contrast, all special relativistic analyses, some dating as far back as the 1950's [2 - 4] and some of newer vintage, [5] predict anharmonic oscillations, even though not all of them agree on the details of these oscillations.

The special relativistic counterpart of our solution is also derivable directly from the Lagrangian: $L(t,x,\dot{t},\dot{x}) = mc\sqrt{\left(1+\omega^2 x^2/c^2\right)\left(c^2\dot{t}^2 - \dot{x}^2\right)}$ , where $\dot{t} = \dfrac{dt}{ds}$, $\dot{x} = \dfrac{dx}{ds}$ and $s$ is an arbitrary monotonically increasing path parameter, eventually identified with the special relativistic proper time (in units of length) of the oscillating mass. This Lagrangian is typical of a relativistic theory, linked with Thirring's name, where a mass and a scalar field interact [6]. We did not find in literature an application of this theory to mass-and-spring systems. An analysis that also considered the implications of a variable mass [7] yielded yet another Lagrangian, of a type known from a scalar field theory linked with Bergmann's name. [6]

## II. KINEMATICAL CONSIDERATIONS

We focus our attention on systems of bodies in motion. The relative space of a body is its geometrical extension to infinity, modelled as the Euclidean space, $\mathbb{R}^3$. (We use the same symbol, $\mathbb{R}^3$, for the corresponding vector space.) Bodies traverse each other's relative space like a plane flying through the earth atmosphere. This causes places from the relative space of one body to coincide with places from the relative space of another body. A place-coincidence is an ephemeral physical event: it happens at one instant when (in principle at least) it could leave marks on bodies, then it vanishes at the next instant. Collectively, the realizable place-coincidences in a multi-body system form the set of physical events that interest us.

In a system with A, B, C … bodies in relative motion, an event $E$ happens at the place $q_{\mathrm{A}}(E) \in \mathbb{R}^3_{\mathrm{A}}$ on the A-body and also at $q_{\mathrm{B}}(E) \in \mathbb{R}^3_{\mathrm{B}}$ on the B-body, at $q_{\mathrm{C}}(E) \in \mathbb{R}^3_{\mathrm{C}}$ on the C-body, and so forth. Events are thus multi-place coincidences. The places that coincide at the event separate thereafter. For example, $q_{\mathrm{A}}(E)$ and $q_{\mathrm{B}}(E)$ separate along a line of relative motion between $\mathbb{R}^3_{\mathrm{A}}$ and $\mathbb{R}^3_{\mathrm{B}}$, the 'AB-line' through $E$. If the direction of the respective AB-lines is the same at all the pairs of coinciding places, and it does not ever change, then the A-body and the B-body moves relative to each other rectilinear and without rotation. The rate of separation between two places that coincide at an event is (in principle) measurable from experiments with two such bodies:

- A light is emitted at some event $E$ and this leaves marks at the place $q_{\mathrm{A}}(E) \in \mathbb{R}^3_{\mathrm{A}}$ and at the place $q_{\mathrm{B}}(E) \in \mathbb{R}^3_{\mathrm{B}}$. The marks keep a record of the two places (one from $\mathbb{R}^3_{\mathrm{A}}$ and the other from $\mathbb{R}^3_{\mathrm{B}}$) that coincide at $E$. The light is detected at some event $D$, and this too leaves marks, at $q_{\mathrm{A}}(D)$ and at $q_{\mathrm{B}}(D)$. Observe how the coincidence of places where an event happens on two bodies in known relative motion identifies this event just as well as the special relativistic time-and-place of the event.
- Now suppose the marks left by the two events, $E$ and $D$, are on the same AB-line. There is ample evidence that light does not propagate instantly. Thus the distance $d_{\mathrm{A}}$ between the places $q_{\mathrm{A}}(D)$ and $q_{\mathrm{A}}(E)$ in $\mathbb{R}^3_{\mathrm{A}}$ is expected to differ from the distance $d_{\mathrm{B}}$ between $q_{\mathrm{B}}(D)$ and $q_{\mathrm{B}}(E)$ in $\mathbb{R}^3_{\mathrm{B}}$. Denote by $k_{\mathrm{AB}}$ the ratio of the shorter to the longer of these distances.
- $k_{\mathrm{AB}}$ measures the (average) rate at which the two places that coincide at either of these events separate thereafter: $0 < k_{\mathrm{AB}} \leq 1$, where $k_{\mathrm{AB}} = 1$ signals no relative motion.

Evidently, the $k_{\mathrm{AB}}$-value is constant when the bodies are inertial. In this sense, a pair of inertial frames could be said to offer a standard of motion for any other body.

The values of $k_{\mathrm{AB}}$, $k_{\mathrm{BC}}$, etc. only partially define the pair-wise relative motions of inertial frames A, B, C, etc. We now envisage unit vectors $\vec{u}_{\mathrm{AB}}$, $\vec{u}_{\mathrm{AC}}$, etc. drawn at $q_{\mathrm{A}}(E)$ on the A-frame in the directions that $q_{\mathrm{A}}(E)$ moves <u>away</u> from $q_{\mathrm{B}}(E)$, $q_{\mathrm{C}}(E)$, etc. after the event $E$. The angle between the unit vectors $\vec{u}_{\mathrm{AB}}$ and $\vec{u}_{\mathrm{AC}}$ is measurable at $q_{\mathrm{A}}(E)$, and so is any other similarly formed angle on any other frame. The $k$-values together with the unit vectors completely determines the motions between all inertial frames.

In practice, tracing the unit vectors is only feasible for coplanar motion, for example, when sheets of paper, $\mathbb{R}^2_{\mathrm{A}}$, $\mathbb{R}^2_{\mathrm{B}}$, $\mathbb{R}^2_{\mathrm{C}}$, etc. slide on each other, on a desktop. Yet, it is still worthwhile to pursue in full generality the logical consequences of tracing and measuring the here defined $k$-values and angles between unit vectors since these pursuits eventually lead to experimentally verifiable conclusions, as illustrated next.

### A. Longitudinal optical Doppler

We analyze here an experiment that produces a longitudinal optical Doppler effect. A series of triggers are placed equally distanced from a light detector on an inertial frame (the A-frame). An emitter on another inertial frame (the B-frame) encounters the triggers as it approaches the detector. Each encounter triggers the emission of a light. The detector senses the light and at each of these events marks the place where it occurred on the B-frame. Let $\lambda_{\mathrm{A}}$ denote the distance between any two consecutive triggers and $\lambda_{\mathrm{B}}$ the distance between any two consecutive marks on the B-frame. The very definition of how the $k_{\mathrm{AB}}$-value is measured leads to the conclusion that: $\lambda_{\mathrm{B}} = k_{\mathrm{AB}}\lambda_{\mathrm{A}}$.

As calculated from a special relativistic analysis of the experiment, the longitudinal Doppler factor expresses the ratio of the frequencies of the emitted light to that of the detected light: $\sqrt{(1+\beta_{\mathrm{AB}})/(1-\beta_{\mathrm{AB}})}$. This is the inverse of the wavelength ratio $\lambda_{\mathrm{B}}/\lambda_{\mathrm{A}}$ that we calculated here. Thus $k_{\mathrm{AB}} = \sqrt{(1-\beta_{\mathrm{AB}})/(1+\beta_{\mathrm{AB}})}$, where $\beta_{\mathrm{AB}}$ is the (fractional) speed of the relative motion between the two inertial frames. (An inertial observer at rest on either frame measures this $\beta_{\mathrm{AB}}$ with his/her frame's synchronized clocks.) Reversing the above relation yields $\beta_{\mathrm{AB}} = \dfrac{1-k_{\mathrm{AB}}{}^2}{1+k_{\mathrm{AB}}{}^2}$, wherefrom $\gamma_{\mathrm{AB}} = \dfrac{1+k_{\mathrm{AB}}{}^2}{2k_{\mathrm{AB}}}$ is the Lorentz factor.

## III. BINATE FRAMES AS REFERENCE SYSTEMS

Two inertial frames, the A-frame and B-frame, whose relative motion is determined by $k_{\mathrm{AB}} = k_{\mathrm{BA}}$ and unit vectors $\left(\vec{u}_{\mathrm{AB}} \in \mathbb{R}^3_{\mathrm{A}}, \vec{u}_{\mathrm{BA}} \in \mathbb{R}^3_{\mathrm{B}}\right)$ forms what we call here a *binate frame*. When clarity is not compromised, we do not list all the parameters of a binate frame: the $k_{\mathrm{AB}}$-frame is shorthand for the $(k_{\mathrm{AB}} = k_{\mathrm{BA}}, \vec{u}_{\mathrm{AB}}, \vec{u}_{\mathrm{BA}})$-frame.

Consider any two events $O$ and $E$. In the $k_{\mathrm{AB}}$-frame, the position vectors: $\vec{r}_{\mathrm{A}} \in \mathbb{R}^3_{\mathrm{A}}$, from $q_{\mathrm{A}}(O)$ to $q_{\mathrm{A}}(E)$, and $\vec{r}_{\mathrm{B}} \in \mathbb{R}^3_{\mathrm{B}}$, from $q_{\mathrm{B}}(O)$ to $q_{\mathrm{B}}(E)$, represent the *displacement* vector $\overrightarrow{OE}$, from $O$ to $E$. The A-plane of this displacement is formed with $\vec{r}_{\mathrm{A}}(E)$ and $\vec{u}_{\mathrm{AB}}$; the B-plane is formed with $\vec{r}_{\mathrm{B}}(E)$ and $\vec{u}_{\mathrm{BA}}$. A diagram like that of Fig. 1 shows the relation between the measurable vectors of the two planes prior to the considered events.

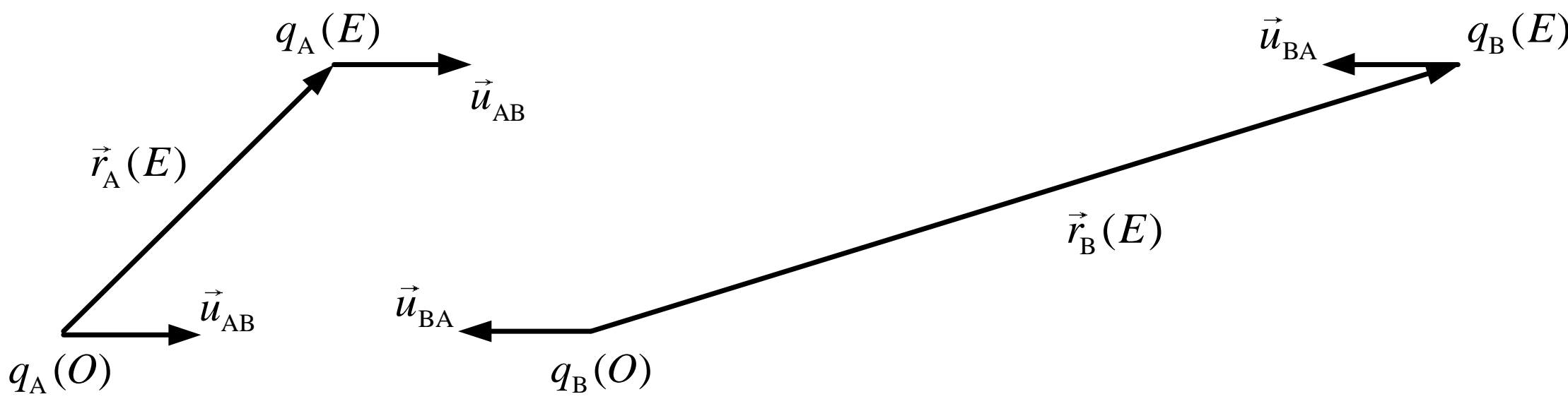

Figure 1. The planes of the $\overrightarrow{OE}$ displacement prior to the considered events

To construct the diagram, we first copy each plane from the respective inertial frame into a common Euclidean $\mathbb{R}^2$, the page of the diagram. We then align $\vec{u}_{\mathrm{AB}}$ with $\vec{u}_{\mathrm{BA}}$ as illustrated. This is possible because the distance between the AB-lines through $q_{\mathrm{A}}(O)$ and $q_{\mathrm{A}}(E)$ on the A-frame is the same as the distance between the AB-lines through $q_{\mathrm{B}}(O)$ and $q_{\mathrm{B}}(E)$ on the B-frame: $|\vec{r}_{\mathrm{A}\perp}| = |\vec{r}_{\mathrm{B}\perp}|$, where $\vec{r}_{\mathrm{A}\perp} = \vec{r}_{\mathrm{A}} - \vec{r}_{\mathrm{A}\parallel}$, $\vec{r}_{\mathrm{A}\parallel} = (\vec{r}_{\mathrm{A}} \bullet \vec{u}_{\mathrm{AB}})\vec{u}_{\mathrm{AB}}$, and there are similarly defined relations for $\vec{r}_{\mathrm{B}\perp}$.

The relative motion between the A-frame and the B-frame is visualized by sliding the two planes along the AB-lines, in the respective directions indicated by the unit vectors $\vec{u}_{\mathrm{AB}}$ and $\vec{u}_{\mathrm{BA}}$. A first two-place coincidence occurs when $q_{\mathrm{A}}(O)$ meets $q_{\mathrm{B}}(O)$. With continued sliding, a second two-place coincidence occurs when $q_{\mathrm{A}}(E)$ meets $q_{\mathrm{B}}(E)$. Observe that $\zeta(\overrightarrow{OE}) = -\vec{r}_{\mathrm{A}}(E) \bullet \vec{u}_{\mathrm{AB}} - \vec{r}_{\mathrm{B}}(E) \bullet \vec{u}_{\mathrm{BA}} > 0$ is *intrinsic* to the $k_{\mathrm{AB}}$-frame since the two scalar products are measurable physical distances (in $\mathbb{R}^3_{\mathrm{A}}$ and respectively $\mathbb{R}^3_{\mathrm{B}}$), thus independent of the coordinates used for these inertial frames. As such, $\zeta(\overrightarrow{OE})$ measures the evolution of the binate frame between the specified events.

Geometrically, the set of realizable events, the multi-place coincidences of a multi-body system, is an affine space. Its points are represented in a $k_{\mathrm{AB}}$-frame by two-place coincidences; its displacements are represented in the same binate frame by the vector-pairs $(\vec{r}_{\mathrm{A}}, \vec{r}_{\mathrm{B}}) \in \mathbb{R}^3_{\mathrm{A}} \times \mathbb{R}^3_{\mathrm{B}}$ that meet the constraints: $|\vec{r}_{\mathrm{A}\perp}| = |\vec{r}_{\mathrm{B}\perp}|$. Such vector pairs form a 4D vector sub-space of $\mathbb{R}^3_{\mathrm{A}} \times \mathbb{R}^3_{\mathrm{B}}$. Evidently, the same affine space could be represented with two-place coincidences from the other binate frames that could be formed (if only momentarily) with pairs of bodies from a multi-body system. We would have more to say about the relations between such representations later on, in Sec VIB.

## IV. LIGHT PROPAGATION

In a $k_{\mathrm{AB}}$-frame, consider again experiments with a light emitted at $O$ and detected at $E$. The diagram of Fig. 2 shows the representative planes of the displacement $\overrightarrow{OE}$ at the detection event $E$, when $q_{\mathrm{A}}(E)$ and $q_{\mathrm{B}}(E)$ coincide. Say we draw similar diagrams at other detection events that see the light from $O$, but retain only the diagrams with (i) the same representative planes as those of $\overrightarrow{OE}$, and (ii) the same distance $\zeta(\overrightarrow{OE})$ between $q_{\mathrm{A}}(O)$ and $q_{\mathrm{B}}(O)$. These diagrams could then be drawn together as in Fig. 2, where all points on the dashed closed curve (here, only partially drawn) represent a detection event just like $E$. We prove next that the locus of the specified detection events is an ellipse.

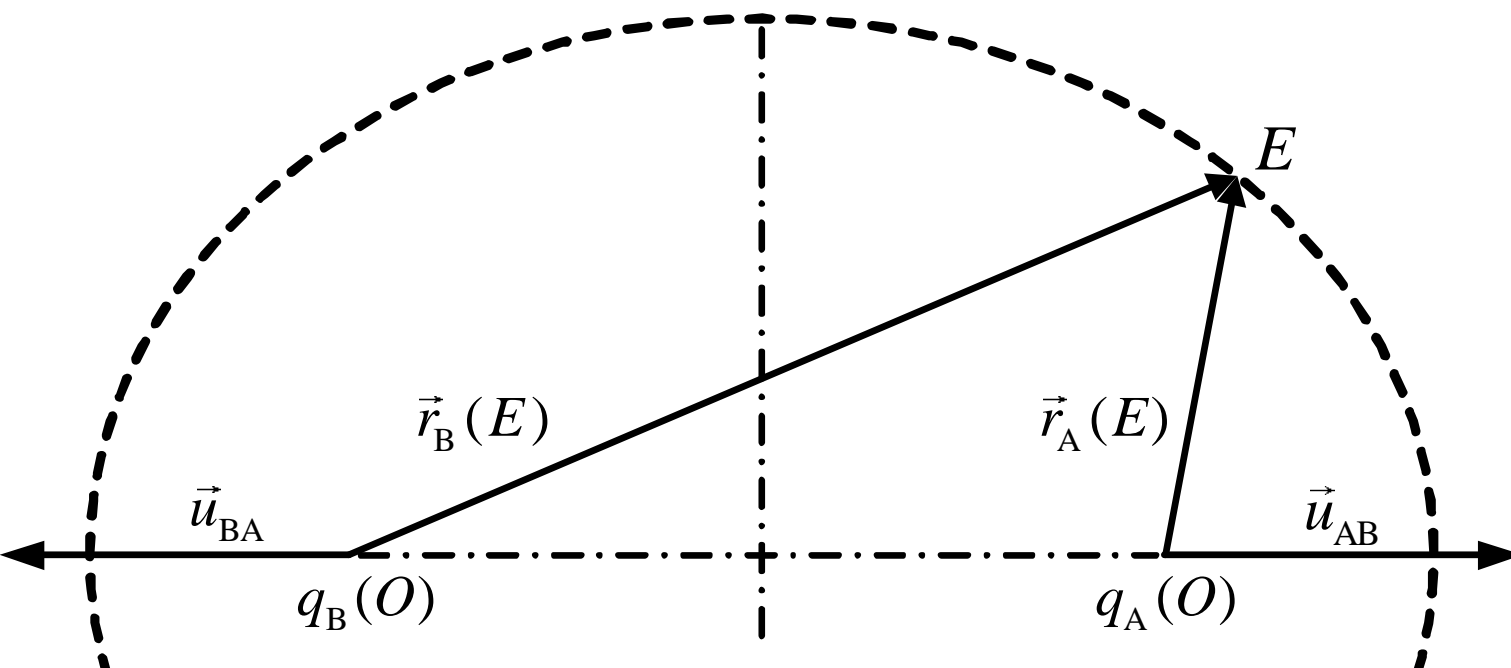

Figure 2. The ellipse associated with emission-detection displacements like $\overrightarrow{OE}$

With electrically neutral bodies of negligibly weak gravitation, there is no physical reason for the displacement $\overrightarrow{OE}$ to be asymmetric relative to the two inertial frames. Thus in Fig. 2, a tangent at $E$ to the locus curve should be equally inclined to the radius vectors, $\vec{r}_{\mathrm{A}}(E)$ and $\vec{r}_{\mathrm{B}}(E)$, otherwise one inertial frame gets preferential treatment. The required equality is the defining property of an ellipse.

We proceed to derive the equation of the ellipse. Recall that the calculations here are based on Euclidean geometry: distances and angles are measured <u>after</u> the experiment, on physical bodies and their relative spaces, that is, on inertial frames that are not anymore in motion relative to each other. (It is irrelevant here whether the so measured distances and angles hold true or not with the frames in relative motion, <u>during</u> the experiment.) The parameters of the ellipse in Fig, 2 are determined from the following considerations:

- The frame's evolution parameter, $\zeta=-\vec{r}_{\mathrm{A}}\bullet\vec{u}_{\mathrm{AB}}-\vec{r}_{\mathrm{B}}\bullet\vec{u}_{\mathrm{BA}}$, is the measured distance between the focuses of the ellipse in Fig. 2.
- The sum $|\vec{r}_{\mathrm{A}}|+|\vec{r}_{\mathrm{B}}|$ of the distances from the place where the detection $E$ occurred to the respective focus on each frame is the length of the major axis of the ellipse.
- The extreme left and right points of the major axis represent two detection events; at each one of these points, the value of $k_{\mathrm{AB}}$ is the ratio of the shorter to the longer of the distances from the point to the two focuses.

Ellipses with eccentricity $\mathrm{e}=(1-k_{\mathrm{AB}})/(1+k_{\mathrm{AB}})$ meet these conditions:

$$\left(1-k_{\mathrm{AB}}\right)\left(|\vec{r}_{\mathrm{A}}|+|\vec{r}_{\mathrm{B}}|\right)=\left(1+k_{\mathrm{AB}}\right)\left(-\vec{r}_{\mathrm{A}}\bullet\vec{u}_{\mathrm{AB}}-\vec{r}_{\mathrm{B}}\bullet\vec{u}_{\mathrm{BA}}\right). \tag{1}$$

We arrived at the light-propagation relation (1) by physical considerations involving a light emitted at $O$ and detectable at all the events on the ellipse of Fig. 2. Other ellipses identical to this ellipse represent the locus of detection events in planes other than those of $\overrightarrow{OE}$, but still for the same value of the frame evolution parameter $\zeta$. We then view the ellipse of Fig. 2 as the longitudinal section through an ellipsoid with a circular cross-section. This ellipsoid is the 2D surface of a light wave front that evolved from $O$ and just arrived at $E$. The evolution of this wave front occurs in the 4D affine space of events, and is represented by a one-parameter family of circular ellipsoids that retain their eccentricity while increasing their size. This family forms a 3D hypersurface embedded in the geometric space events. It is the binate frame counterpart of the special relativistic light-cone. Only events on this hypersurface detect the light emitted at $O$.

Observe that the evolving family of ellipsoids, like the special relativistic light-cone, does not fully determine the light-ray $\widetilde{OE}$, that is, the sequence of events $D_\zeta$ that could detect the light emitted at $O$ before it is extinguished at $E$. For such determination, more is needed, as explained next.

## A. Light rays

The light wave-front relation (1) could be split into a pair of relations that respect the AB-symmetry of the binate frame. One relation describes the trace of the light-ray $\widetilde{OE}$ through the relative space of the A-frame, the other, the trace through the B-frame. There is ample evidence that (in negligible gravitational fields) light propagates rectilinearly unless specific physical factors in the environment prevent this. Thus the light-ray $\widetilde{OE}$ traces the line between $q_\text{A}(O)$ and $q_\text{A}(E)$ on the A-frame and the line between $q_\text{B}(O)$ and $q_\text{B}(E)$ on the B-frame. Only devices placed along these lines could detect the light.

The parametric equations of the light-ray $\widetilde{OE}$ are: $\vec{r}_\text{A} = \rho_\text{A}(\zeta)\vec{1}_{r\text{A}}$ and $\vec{r}_\text{B} = \rho_\text{B}(\zeta)\vec{1}_{r\text{B}}$, where $\vec{1}_{r\text{A}}$ and $\vec{1}_{r\text{B}}$ are unit vectors along the $\overrightarrow{OE}$'s projection on the respective inertial frames. To find $\rho_\text{A}(\zeta)$ and $\rho_\text{B}(\zeta)$, substitute $\zeta(\overrightarrow{OD_\zeta})$ for $\left[-\vec{r}_\text{A}(D_\zeta)\bullet\vec{u}_\text{AB} - \vec{r}_\text{B}(D_\zeta)\bullet\vec{u}_\text{BA}\right]$ in (1) then complement this with a second equation: $\rho_\text{A}\sin(\alpha_\text{A}) = \rho_\text{B}\sin(\alpha_\text{B})$, where $\alpha_\text{A}, \alpha_\text{B} \le \pi$ are the angles between $\vec{r}_\text{A}$ and $\vec{u}_\text{AB}$, respectively $\vec{r}_\text{B}$ and $\vec{u}_\text{BA}$. These yield:

$$\rho_\text{A} = \frac{1+k_\text{AB}}{1-k_\text{AB}} \frac{\zeta \sin(\alpha_\text{B})}{\sin(\alpha_\text{A})+\sin(\alpha_\text{B})} \quad \text{and} \quad \rho_\text{B} = \frac{1+k_\text{AB}}{1-k_\text{AB}} \frac{\zeta \sin(\alpha_\text{A})}{\sin(\alpha_\text{A})+\sin(\alpha_\text{B})}. \qquad (1a)$$

Incidentally, substitutions from (1a) into (1) yield: $\cot(\tfrac{1}{2}\alpha_\text{A})\cot(\tfrac{1}{2}\alpha_\text{B}) = k_\text{AB}$. In a slightly different form, this relation appeared in a more than a century old analysis of stellar aberration. [8] The transversal optical Doppler effect is easily predicted from this old relation, now applied to a situation where $E$ in Fig. 2 is the emission event and $O$ is the detection event that marks the inertial frames at $O_\text{A}$ and respectively $O_\text{B}$.

## B. Binate-frame analysis of a Michelson-Morley type of experiment

Consider a cart in uniform motion on a linear track. An apparatus on its platform has: (i) an arm of length $L$, inclined with respect to track; (ii) a source and a detector of light at one end of the arm; (iii) a mirror at the other end. An experiment is performed where the source emits a light that the mirror reflects it back to the source.

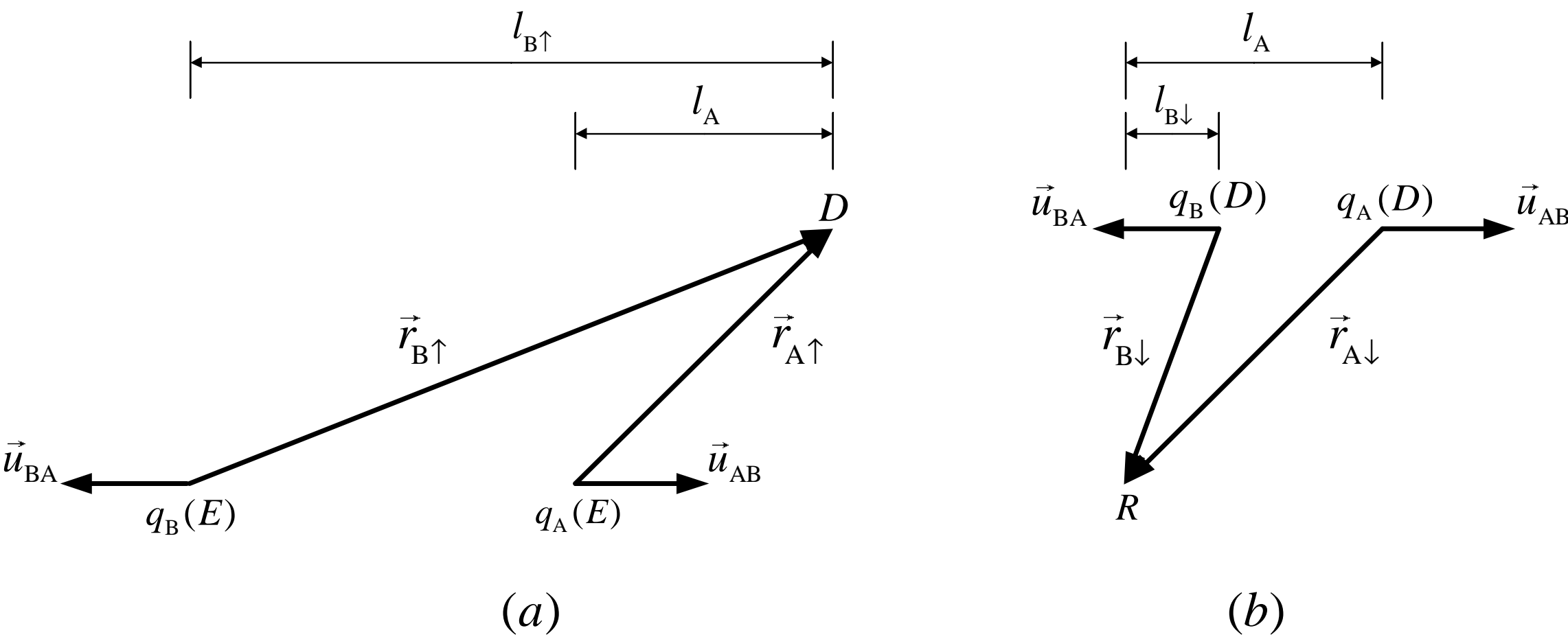

Figure 3. Displacements $\overrightarrow{ED}$ and $\overrightarrow{DR}$

In a $k_{AB}$ -frame adapted to the experiment, the cart with the apparatus is the A-frame and the track is the B-frame. The source emits the light at $E$, which marks the track at $q_B(E)$. The mirror detects the light at $D$ and reflects it back to the source. This detects the reflected light at $R$ and marks the track at $q_B(R)$. The planes of the displacements $\overrightarrow{ED}$ and $\overrightarrow{DR}$ are illustrated with the diagrams of Fig. 3:

- In Fig. (3a), $\overrightarrow{ED}$ is shown at the first detection $D$, when $q_A(D)$ meets $q_B(D)$. The geometry of the figure provides the relations: $\left|\vec{r}_{A\uparrow}\right| = L$, $\left|\vec{r}_{B\uparrow}\right| = \sqrt{{l_{B\uparrow}}^2 + L^2 - {l_A}^2}$, and $-\vec{r}_{A\uparrow} \bullet \vec{u}_{AB} - \vec{r}_{B\uparrow} \bullet \vec{u}_{BA} = l_{B\uparrow} - l_A$. We substitute them in (1) and solve it for $l_{B\uparrow}$. This yields: $l_{B\uparrow} = \dfrac{1 + {k_{AB}}^2}{2k_{AB}} l_A + \dfrac{1 - {k_{AB}}^2}{2k_{AB}} L$.
- In Fig. 3(b), $\overrightarrow{DR}$ is shown at the second detection $R$, when $q_A(R)$ meets $q_B(R)$. The geometry of the figure provides the relations: $\left|\vec{r}_{A\downarrow}\right| = L$, $\left|\vec{r}_{B\downarrow}\right| = \sqrt{{l_{B\downarrow}}^2 + L^2 - {l_A}^2}$ and $-\vec{r}_{A\downarrow} \bullet \vec{u}_{AB} - \vec{r}_{B\downarrow} \bullet \vec{u}_{BA} = l_A - l_{B\downarrow}$. We substitute these in (1) then solve it for $l_{B\downarrow}$. This yields: $l_{B\downarrow} = \dfrac{1 + {k_{AB}}^2}{2k_{AB}} l_A - \dfrac{1 - {k_{AB}}^2}{2k_{AB}} L$.

(To clearly separate from each other all the vectors presented in the diagram, we used a binate frame with $k_{\mathrm{AB}} = .5$. The corresponding Lorentz factor is $\gamma_{\mathrm{AB}} = 1.25$.)

The distance on the track between the mark placed at the emission of the light and the mark placed when the source detected the reflection from the mirror is $d = l_{\mathrm{B}\uparrow} - l_{\mathrm{B}\downarrow}$. We substitute here our earlier results to get $d = \left(k_{\mathrm{AB}}^{-1} - k_{\mathrm{AB}}\right)L$. This does not depend on the arm's inclination. Thus if we attach a second arm of the same length $L$ but now inclined at $90^o$ with respect to the first, as in the famous Michelson-Morley experiment, the source would detect the reflected rays from the two mirrors simultaneously!

## V. PARAMETERIZATION WITH CARTESIAN POSITION-COORDINATES

A displacement $\overrightarrow{OE}$ represented in a $k_{\mathrm{AB}}$-frame by the pair of radius vectors $\left(\vec{r}_{\mathrm{A}}, \vec{r}_{\mathrm{B}}\right)$ from $O$ to $E$ could be used to generate a parameterization of events that preserves the AB-symmetry of the frame. It suffice to select: An origin of coordinates, at $q_{\mathrm{A}}(O)$ and respectively $q_{\mathrm{B}}(O)$. An orthonormal basis for the vectors on the A-frame: $\hat{1}_{x\mathrm{A}} = \vec{u}_{\mathrm{AB}}$, $\hat{1}_{y\mathrm{A}} = \vec{r}_{\mathrm{A}} \times \vec{u}_{\mathrm{AB}} / \left|\vec{r}_{\mathrm{A}} \times \vec{u}_{\mathrm{AB}}\right|$, and $\hat{1}_{z\mathrm{A}} = \hat{1}_{x\mathrm{A}} \times \hat{1}_{y\mathrm{A}}$. A reciprocal basis on B-frame: $\hat{1}_{x\mathrm{B}} = \vec{u}_{\mathrm{BA}}$, $\hat{1}_{y\mathrm{B}} = \vec{r}_{\mathrm{B}} \times \vec{u}_{\mathrm{BA}} / \left|\vec{r}_{\mathrm{B}} \times \vec{u}_{\mathrm{BA}}\right|$, and $\hat{1}_{z\mathrm{B}} = \hat{1}_{x\mathrm{B}} \times \hat{1}_{y\mathrm{B}}$. The distance from the AB-line through $O$ to the AB-line through $E$ could serve as the unit-length since it is the same on both bodies. The coordinates of other events $\overline{E}$ then obtain as usual, by projecting the position vectors $\vec{r}_{\mathrm{A}}(\overline{E})$ and $\vec{r}_{\mathrm{B}}(\overline{E})$ on the respective unit vectors. Since the relative motion of the inertial frames enforces: $y_{\mathrm{A}}(\overline{E}) = -y_{\mathrm{B}}(\overline{E})$ and $z_{\mathrm{A}}(\overline{E}) = z_{\mathrm{B}}(\overline{E})$, only four of the so obtained six coordinates of an event are independent.

### A. Coordinate expressions for the light-propagation relation

Inserting in (1) the coordinates of an event that detects a light from $O$ yields the following equation, where to reflect the intrinsic symmetry of the AB-pair of reference bodies, we did not use for the parenthesis on the right hand side either one of its two equivalent forms: $\left(-y_{\mathrm{A}}^2 - z_{\mathrm{A}}^2\right)$ or $\left(-y_{\mathrm{B}}^2 - z_{\mathrm{B}}^2\right)$.

$$4k_{\mathrm{AB}}^{2}\left(x_{\mathrm{A}}^{2} + \frac{1 + k_{\mathrm{AB}}^{2}}{k_{\mathrm{AB}}} x_{\mathrm{A}} x_{\mathrm{B}} + x_{\mathrm{B}}^{2}\right) = \left(1 - k_{\mathrm{AB}}^{2}\right)^{2}\left(z_{\mathrm{A}} z_{\mathrm{B}} - y_{\mathrm{A}} y_{\mathrm{B}}\right). \qquad (1b)$$

Note that regardless of the sign on the right-hand side of (1), the same (1b) obtains. But only those solutions to (1b) that meet $\zeta = \left(-x_{\mathrm{A}} - x_{\mathrm{B}}\right) = \left(-\vec{r}_{\mathrm{A}} \bullet \vec{u}_{\mathrm{AB}} - \vec{r}_{\mathrm{B}} \bullet \vec{u}_{\mathrm{BA}}\right) > 0$ are physically acceptable. The emission of a light precedes its detection!

The vector nature of a differential displacement is best highlighted with the familiar ‘ dx ’ notation: $dx^1 = dx_{\mathrm{A}}$, $dx^2 = dx_{\mathrm{B}}$, $dx^3 = dy_{\mathrm{A}} = -dy_{\mathrm{B}}$, and $dx^4 = dz_{\mathrm{A}} = dz_{\mathrm{B}}$. Thus, with usual summation convention on repeated indices, (1b) becomes $\eta_{\mu\nu}(k_{\mathrm{AB}})dx^\mu dx^\nu = 0$, or its equivalent $dx_\mu dx^\mu = 0$, with $dx_\mu = \eta_{\mu\nu}dx^\nu$ the covariant form of the displacement; physically meaningful displacements meet $d\zeta = -dx^1 - dx^2 > 0$. The non-null entries of $\eta_{\mu\nu}(k_{\mathrm{AB}})$ are: $\eta_{11} = \eta_{22} = \left(\dfrac{2k_{\mathrm{AB}}}{1-k_{\mathrm{AB}}{}^2}\right)^2$, $\eta_{12} = \eta_{21} = \left(\dfrac{2k_{\mathrm{AB}}}{1-k_{\mathrm{AB}}{}^2}\right)\left(\dfrac{1+k_{\mathrm{AB}}{}^2}{1-k_{\mathrm{AB}}{}^2}\right)$ and $\eta_{33} = \eta_{44} = -1$.

## VI. PARTICLES AND THEIR PATH INVARIANT

Consider anew the diagram of Fig. 2 for a light emitted at $O$ and detected nearby at $E$. There is ample evidence that a massive particle could be present at only one of these two events: In a $k_{\mathrm{AB}}$-frame, the particle was either at the place $q_{\mathrm{A}}(O)$ before it coincided with the place $q_{\mathrm{B}}(O)$, and the emission event $O$ happened. Or it was at $q_{\mathrm{A}}(E)$ after it coincided with $q_{\mathrm{B}}(E)$, and the detection event $E$ happened. When these considerations are applied in conjunction with (1b), they show that $ds^2 = \eta_{\mu\nu}(k_{\mathrm{AB}})dx^\mu dx^\nu > 0$ between near events attended by the particle. Thus the so defined differential *separation* interval, $ds$, is indeed real valued along the particle’s path. In terms of canonical coordinates:

$$ds^2 = \left(\frac{2k_{\mathrm{AB}}}{1-k_{\mathrm{AB}}{}^2}\right)^2 (dx_{\mathrm{A}}{}^2 + \frac{1+k_{\mathrm{AB}}{}^2}{k_{\mathrm{AB}}} dx_{\mathrm{A}} dx_{\mathrm{B}} + dx_{\mathrm{B}}{}^2) + dy_{\mathrm{A}}\, dy_{\mathrm{B}} - dz_{\mathrm{A}}\, dz_{\mathrm{B}}\,. \qquad (2)$$

In neighborhoods of finite extent, the path of a particle present at the events $E_1$ and $E_2$ need not be the same as the path of another particle also present at these events. This is evident from observing that not all the particles trace the same curves on the inertial frames of the $k_{\mathrm{AB}}$-frame. The (finite) separation between the specified two events along the path $x^\mu(\xi)$ of a particle obtains as the path integral:

$$s = \int_{\xi(E_1)}^{\xi(E_2)} \sqrt{\eta_{\mu\nu}(k_{\mathrm{AB}})\, x'^\mu(\xi)\, x'^\nu(\xi)}\, d\xi > 0\,. \qquad (3)$$

The integral reaches the greatest value when: $\dfrac{d}{d\xi}\left[\dfrac{\partial}{\partial x'^\sigma}\sqrt{\eta_{\mu\nu}(k_{\mathrm{AB}})x'^\mu(\xi)x'^\nu(\xi)}\right] = 0$, that is, along a path that projects straight lines on the inertial frames that form the $k_{\mathrm{AB}}$-frame. The maximal value does not depend on the choice of the path parameter $\xi$, though the rates of change $x'^\mu(\xi)$, which are constant on this ‘maximal’ path, evidently depend of this parameter.

Say a particle on such a maximal path marks at three events the two inertial frames of some $k_{\mathrm{AB}}$-frame. Then on each inertial frame, the three marks would be aligned and the ratio of distances between any pair of marks would be the same on both frames. This is just how classical mechanics expects a particle to move when it is free from external influences. It justifies the following kinematic definition of a *free* particle: it is a particle that moves along the maximizing path between any pair of events that it attends.

It is obvious from (2) that in a $k_{\mathrm{AB}}$-frame adapted to its motion, a free particle at rest at the origin of A-frame coordinates and present at $x_{\mathrm{B}}=0$ could attend only the events with $dx_{\mathrm{B}}=-\frac{1-k_{\mathrm{AB}}{}^{2}}{2k_{\mathrm{AB}}}ds\leq 0$. To compare with special relativity, substitute $k_{\mathrm{AB}}$ in terms of $\beta_{\mathrm{AB}}$. This yields $\beta_{\mathrm{AB}}=\frac{-dx_{\mathrm{B}}}{\gamma_{\mathrm{AB}}ds}$, which is the special relativistic speed of the particle as measured by an observer at rest on the B-frame.

When two free particles collide, they may coalesce and thereafter move together. The free particles are initially at rest on their respective inertial frames, and these form a $k_{\mathrm{AB}}$-frame adapted to the relative motion between the two particles. With identical particles, the coalesced particle would move symmetric relative to the inertial frames, and from (2), we get $\frac{dx_{\mathrm{A}}}{ds}=\frac{dx_{\mathrm{B}}}{ds}=-\frac{1-k_{\mathrm{AB}}}{2\sqrt{k_{\mathrm{AB}}}}$. With particles of different mass, this symmetry is broken: The coalesced particle would be at rest on an inertial frame that forms a $k_{\mathrm{PA}}$-frame with the A-frame, and another $k_{\mathrm{PB}}$-frame with the B-frame. From (2), we get: $\frac{dx_{\mathrm{A}}}{ds}=-\frac{1-k_{\mathrm{PA}}{}^{2}}{2k_{\mathrm{PA}}}$ and $\frac{dx_{\mathrm{B}}}{ds}=-\frac{1-k_{\mathrm{PB}}{}^{2}}{2k_{\mathrm{PB}}}$. The $k$-values so defined meet $k_{\mathrm{PA}}k_{\mathrm{PB}}=k_{\mathrm{AB}}$. If we now substitute the three $k$ in terms of the respective $\beta$, we get the special relativistic relation between collinear velocities.

### A. Invariance of the separation interval under a change of parameterization

The freedom to choose a displacement that generates the canonical parameterization of events in a $k_{\mathrm{AB}}$-frame leads to a group of coordinate-transformations composed of a four-parameter translation sub-group and a one-parameter sub-group of rotations around the AB-line through the origins of the coordinates. The integrant in (3) depends only on the differences between the coordinates, not on the coordinates themselves, and thus it is invariant under the translation sub-group. It is also invariant under the rotation sub-group since the two equivalent forms of (2) below involve only $\mathbb{R}^{3}$ scalar products:

$$ds^2 = \left( \frac{1+k_{\mathrm{AB}}{}^2}{1-k_{\mathrm{AB}}{}^2} d\vec{r}_{\mathrm{A}} \bullet \vec{u}_{\mathrm{AB}} + \frac{2k_{\mathrm{AB}}}{1-k_{\mathrm{AB}}{}^2} d\vec{r}_{\mathrm{B}} \bullet \vec{u}_{\mathrm{BA}} \right)^2 - \left| d\vec{r}_{\mathrm{A}} \right|^2 , \qquad (2a)$$

$$ds^2 = \left( \frac{2k_{\mathrm{AB}}}{1-k_{\mathrm{AB}}{}^2} d\vec{r}_{\mathrm{A}} \bullet \vec{u}_{\mathrm{AB}} + \frac{1+k_{\mathrm{AB}}{}^2}{1-k_{\mathrm{AB}}{}^2} d\vec{r}_{\mathrm{B}} \bullet \vec{u}_{\mathrm{BA}} \right)^2 - \left| d\vec{r}_{\mathrm{B}} \right|^2 . \qquad (2b)$$

We conclude that the separation interval between a pair of events along the path of a particle in a specified binate frame is invariant under all changes of the displacement that generates the canonical parameterization of events.

The relativistic counterpart of (2a) and (2b) is $ds^2 = (ct_{\mathrm{A}})^2 - |d\vec{r}_{\mathrm{A}}|^2 = (ct_{\mathrm{B}})^2 - |d\vec{r}_{\mathrm{B}}|^2$, where $t_{\mathrm{A}}$ and $t_{\mathrm{B}}$ are the time-coordinates of observers located on the respective inertial frames of the $k_{\mathrm{AB}}$-frame, and $c$ is the speed of light as measured by either of them. We now substitute $x_{\mathrm{A}} = \vec{r}_{\mathrm{A}} \bullet \vec{u}_{\mathrm{AB}}$ and $x_{\mathrm{B}} = \vec{r}_{\mathrm{A}} \bullet \vec{u}_{\mathrm{BA}}$ in the above relations for $ds^2$ to get:

$$ct_{\mathrm{A}} = -\frac{1+k_{\mathrm{AB}}{}^2}{1-k_{\mathrm{AB}}{}^2} x_{\mathrm{A}} - \frac{2k_{\mathrm{AB}}}{1-k_{\mathrm{AB}}{}^2} x_{\mathrm{B}} \quad \text{and} \quad ct_{\mathrm{B}} = -\frac{2k_{\mathrm{AB}}}{1-k_{\mathrm{AB}}{}^2} x_{\mathrm{A}} - \frac{1+k_{\mathrm{AB}}{}^2}{1-k_{\mathrm{AB}}{}^2} x_{\mathrm{B}} . \qquad (4)$$

(The minus signs ensure that the calculated times are positive when $\zeta = -x_{\mathrm{A}} - x_{\mathrm{B}} > 0$.)

Say we view the B-frame as 'stationary' and the A-frame as 'moving'. Then in the standard configuration of special relativity, the $x'$-axis on the A-frame is oriented away from the B-frame. This is the same orientation as our unit vector $\vec{u}_{\mathrm{AB}}$. In contrast, the $x$-axis on the B-frame is oriented toward the A-frame. This is opposite to the orientation of our $\vec{u}_{\mathrm{BA}}$. Accordingly, we substitute in (4): $x_{\mathrm{A}} = x'$, $x_{\mathrm{B}} = -x$, and $k_{\mathrm{AB}}$ in terms of $\beta_{\mathrm{AB}}$, to get the Lorentz transformations, which are now reformatted as definitions of the two time-coordinates, $t' = t_{\mathrm{A}}$ and $t = t_{\mathrm{B}}$, used by special relativity.

## B. Invariance of the separation interval under a change of representation

Let $\overrightarrow{E_1 E_2}$ denote the *displacement* from an event $E_1$ to an event $E_2$. It is represented in a $k_{\mathrm{AB}}$-frame by the vector-pair: $\left( \overrightarrow{q_{\mathrm{A}}(E_1) q_{\mathrm{A}}(E_2)},\ \overrightarrow{q_{\mathrm{B}}(E_1) q_{\mathrm{B}}(E_2)} \right)$. We mentioned at the conclusion of Sec. III that such vector-pairs form a 4D vector sub-space of $\mathbb{R}^3_{\mathrm{A}} \times \mathbb{R}^3_{\mathrm{B}}$. A basis for this vector sub-space obtains with four independent light-ray displacements, for example, the following vector-pairs here presented in a canonical parameterization:

$$\left(\vec{e}_{\mathrm{A1}},\vec{e}_{\mathrm{B1}}\right)=\left(\begin{bmatrix}k_{\mathrm{AB}}\\0\\0\end{bmatrix},\begin{bmatrix}-1\\0\\0\end{bmatrix}\right),\ \left(\vec{e}_{\mathrm{A2}},\vec{e}_{\mathrm{B2}}\right)=\left(\begin{bmatrix}0\\1\\0\end{bmatrix},\begin{bmatrix}-\dfrac{1-k_{\mathrm{AB}}{}^{2}}{2k_{\mathrm{AB}}}\\-1\\0\end{bmatrix}\right),\ \left(\vec{e}_{\mathrm{A3}},\vec{e}_{\mathrm{B3}}\right)=\left(\begin{bmatrix}-1\\0\\0\end{bmatrix},\begin{bmatrix}k_{\mathrm{AB}}\\0\\0\end{bmatrix}\right),$$

and $\left(\vec{e}_{\mathrm{A4}},\vec{e}_{\mathrm{B4}}\right)=\left(\begin{bmatrix}-\dfrac{1-k_{\mathrm{AB}}{}^{2}}{2k_{\mathrm{AB}}}\\0\\1\end{bmatrix},\begin{bmatrix}0\\0\\1\end{bmatrix}\right)$.

A differential displacement is a weighted sum of these four light-ray vector-pairs: $d\vec{r}_{\mathrm{A}}=\sum_{i=1}^{4}d\alpha_i\,\vec{e}_{\mathrm{A}i}$ and $d\vec{r}_{\mathrm{B}}=\sum_{i=1}^{4}d\alpha_i\,\vec{e}_{\mathrm{B}i}$. When changing from the $\left(k_{\mathrm{AB}},\vec{u}_{\mathrm{AB}}\right)$-frame to say the $\left(k_{\mathrm{CB}},\vec{u}_{\mathrm{CB}}\right)$-frame, the weighting factors $d\alpha_i$ retain their values since they are absolute scalars, but the $e$-basis above changes to the corresponding $\varepsilon$-basis of the new frame, as explained next.

Say the selected displacement that generated the canonical parameterisation of the two inertial frames is represented by $\left(\vec{r}_{\mathrm{A}},\vec{r}_{\mathrm{B}}\right)$ in the $\left(k_{\mathrm{AB}},\vec{u}_{\mathrm{AB}}\right)$-frame and by $\left(\vec{r}_{\mathrm{C}},\vec{r}_{\mathrm{B}}\right)$, in the $\left(k_{\mathrm{CB}},\vec{u}_{\mathrm{CB}}\right)$-frame. The two binate frames have in common the B-frame; we call it the *pivot* of this frame change. Observe that $\mathbb{R}^3_{\mathrm{B}}$ has now two vector bases:

- From the $\left(k_{\mathrm{AB}},\vec{u}_{\mathrm{AB}}\right)$-frame: $\hat{1}_{x\mathrm{B}}=\vec{u}_{\mathrm{BA}}$, $\hat{1}_{y\mathrm{B}}=\vec{r}_{\mathrm{B}}\times\vec{u}_{\mathrm{BA}}/\left|\vec{r}_{\mathrm{B}}\times\vec{u}_{\mathrm{BA}}\right|$, and $\hat{1}_{z\mathrm{B}}=\hat{1}_{x\mathrm{B}}\times\hat{1}_{y\mathrm{B}}$;
- From the $\left(k_{\mathrm{CB}},\vec{u}_{\mathrm{CB}}\right)$-frame: $\tilde{1}_{x\mathrm{B}}=\vec{u}_{\mathrm{BC}}$, $\tilde{1}_{y\mathrm{B}}=\vec{r}_{\mathrm{B}}\times\vec{u}_{\mathrm{BC}}/\left|\vec{r}_{\mathrm{B}}\times\vec{u}_{\mathrm{BC}}\right|$, and $\tilde{1}_{z\mathrm{B}}=\tilde{1}_{x\mathrm{B}}\times\tilde{1}_{y\mathrm{B}}$.

The basis vectors $\vec{\varepsilon}_{\mathrm{B}i}$ are easily calculated as the projections of the vectors $\vec{e}_{\mathrm{B}i}$ on the new coordinate axes. From (2), we then get an equation for $\varepsilon_{\mathrm{C}i}{}^{1}$:

$$\frac{2k_{\mathrm{AB}}}{1-k_{\mathrm{AB}}{}^{2}}e_{\mathrm{Ai}}{}^{1}+\frac{1+k_{\mathrm{AB}}{}^{2}}{1-k_{\mathrm{AB}}{}^{2}}e_{\mathrm{Bi}}{}^{1}=\frac{2k_{\mathrm{CB}}}{1-k_{\mathrm{CB}}{}^{2}}\varepsilon_{\mathrm{Ci}}{}^{1}+\frac{1+k_{\mathrm{CB}}{}^{2}}{1-k_{\mathrm{CB}}{}^{2}}\varepsilon_{\mathrm{Bi}}{}^{1}. \tag{5}$$

Finally, the other two components of $\varepsilon_{\mathrm{C}i}$ obtain from: $\varepsilon_{\mathrm{Bi}}{}^{2}+\varepsilon_{\mathrm{Ci}}{}^{2}=0$ and $\varepsilon_{\mathrm{Bi}}{}^{3}=\varepsilon_{\mathrm{Ci}}{}^{3}$.

Relations of the same form as (5) hold in general, not just for light rays, since a differential displacement along the path of a particle is a linear combination of the four basis vectors. Thus a differential displacement that is represented in the $k_{\mathrm{AB}}$-frame by the vector-pairs $\left(d\vec{r}_{\mathrm{A}},d\vec{r}_{\mathrm{B}}\right)$ and in the $k_{\mathrm{CB}}$-frame by $\left(d\vec{r}_{\mathrm{C}},d\vec{r}_{\mathrm{B}}\right)$ meets:

$$\frac{2k_{\mathrm{AB}}}{1-k_{\mathrm{AB}}{}^{2}}d\vec{r}_{\mathrm{A}}\bullet\vec{u}_{\mathrm{AB}}+\frac{1+k_{\mathrm{AB}}{}^{2}}{1-k_{\mathrm{AB}}{}^{2}}d\vec{r}_{\mathrm{B}}\bullet\vec{u}_{\mathrm{BA}}=\frac{2k_{\mathrm{CB}}}{1-k_{\mathrm{CB}}{}^{2}}d\vec{r}_{\mathrm{C}}\bullet\vec{u}_{\mathrm{CB}}+\frac{1+k_{\mathrm{CB}}{}^{2}}{1-k_{\mathrm{CB}}{}^{2}}d\vec{r}_{\mathrm{B}}\bullet\vec{u}_{\mathrm{BC}}. \tag{6}$$

Note that $|d\vec{r}_{\text{B}}|$ has the same value in the two binate-frames. But then (2b) and (6) show that also the value of the finite separation along a defined path between any two specified events does not depend on the choice of the binate frame used to represent the considered path and events.

We conclude that separation interval between any pairs of events along the path of a particle retains its value not only when changing the parameterization of a binate frame, but also when changing the binate frame itself. All matters regarding the relation between the Lorentz group of transformations between the spacetimes of inertial observers and the group of transformations between the representations of events in different binate frames are deferred to a later paper.

## C. On the radioactive decay of muons in flight

Radioactive particles decay over time in a random manner. When measured in the laboratory, the time-rate of decay $dN/dt$ depends on the number $N$ of particles at the time of measurement. This leads to the prediction that $N = N_0 e^{-t/T}$, where $N_0$ is the number of particles at the start of a measured interval of duration $t$, $N$ is the number of those that survived at the end of this interval, and $T$ is their *average lifetime*, the constant of best exponential fit to the experimental curve obtained in the laboratory.

Applying the above formula to the cosmic–ray muons that entered the atmosphere, with $N_0$ measured at the top of a mountain and $N$ measured at the sea level, leads to predictions that underestimate the survival proportion of these muons. Special relativistic considerations of time dilation or length contraction require that the proper time $\tau$ of the decaying particle be used instead of the laboratory time, $t$. Indeed, $N = N_0 e^{-\tau/T}$ yields a much better estimate.

Observer-dependent notions of time are not available when events are represented in a binate frame. The rate of decay is a physical property of the particle: it does not depend on the frame used for its measurement. Since the separation interval $s$ is independent of the frame, it could be used as the parameter of the decay curve: $N = N_0 e^{-s/S}$, where $S$ is the *average lifespan*, the experimental constant of best exponential fit.

The only relevant motion is between the muons captured at the mountaintop and the ones allowed to continue their flight. A $k_{\text{AB}}$-frame adapted to this motion is formed with the relative space of the muon in flight (A-frame) and the relative space of the mountain (B-frame). The muon arrives at the mountaintop at the initial event $O$ and reaches the sea level at the event $E$. The finite interval from $O$ to $E$ along the path of a muon in flight is: $s_{\text{A}} = -\dfrac{2k_{\text{AB}}}{1-k_{\text{AB}}{}^2} x_{\text{B}}(E)$. It obtains from (2) and (3) with $|x_{\text{B}}(E)|$ the mountain height.

The special relativistic counterparts of (2a), (2b) and (4) show that the interval $s$ here is the estimated proper time $\tau$ in units of lengths: $s = c\tau$. Then $S = cT$ is the constant of best fit measured with the laboratory clocks but presented in units of length. (One could obtain $S$ directly, by counting the number of surviving muons that pass-by at different altitudes, but we did not find in the literature reliable records of such counts.) Substitution in $N = N_0 e^{-s_A/S}$ yields the experimentally verified prediction of special relativity, though here neither the notion of time dilation not that of length contraction plays any role.

## VII. KINEMATICS OF A THREE-BODY SYSTEM

We analyze next a situation that arises typically in a three-body system when each body moves in a different direction. It offers the reader on one hand, an opportunity to check that everything explained so far is well understood, and on the other hand, a most direct explanation for the Wigner angle, the kinematic effect behind Thomas precession.

Two blank pages move uniformly, without rotations, on a desktop D: the E-page eastward, the N-page northward. A device on the desktop emits a light signal at $O$, and this event marks the E- and N-page, at $q_{\mathrm{E}}(O)$ and $q_{\mathrm{N}}(O)$, respectively. Eastward and northward from the emitter, at equal distances from it, are two detectors. The eastward detector senses the light at the event $E$ and puts marks on the two pages, at $q_{\mathrm{E}}(E)$ and $q_{\mathrm{N}}(E)$. The northward detector senses the light at $N$ and puts marks at $q_{\mathrm{E}}(N)$ and $q_{\mathrm{N}}(N)$. The marks are the outcome of the experiment. They provide a record of the relevant displacements: $\overrightarrow{OE}$, $\overrightarrow{ON}$ and $\overrightarrow{EN}$.

After the experiment, we find marks at $q_{\mathrm{E}}(O)$, $q_{\mathrm{E}}(E)$ and $q_{\mathrm{E}}(N)$ on the E-page. Let $\mathrm{E}_{\Delta}$ be the triangle with corners at these places. We also find marks at $q_{\mathrm{N}}(O)$, $q_{\mathrm{N}}(E)$ and $q_{\mathrm{N}}(N)$ on the N-page; these form the $\mathrm{N}_{\Delta}$ triangle. The emitter and the two detectors on the desktop form the triangle $\mathrm{D}_{\Delta}$ with corners at $q_D(O)$, $q_D(E)$, and $q_D(N)$.

We copy the triangles $\mathrm{E}_{\Delta}$ and $\mathrm{N}_{\Delta}$ to the desktop and align them as in Fig. 4, along with the unit vectors $\vec{u}_{\mathrm{DE}}$ and $\vec{u}_{\mathrm{DN}}$. The resulting diagram shows the relation between the triangles prior to the emission event: $E_{\Delta}$ is westward and $N_{\Delta}$ is southward from $D_{\Delta}$. (In Fig. 4, the shapes of the triangles are as calculated for $k_{\mathrm{DE}} = .6$ and $k_{\mathrm{DN}} = .8$.)

The E-page and the N-page move uniformly relative to the desktop. We could then form two binate frames:

- The $k_{\mathrm{ED}}$-frame is adapted to the motion of the E-page: $\vec{u}_{\mathrm{ED}}$ is eastward. After the experiment, we could measure the distance $|\vec{r}_{\mathrm{E}}(E)|$ between the marks at $q_{\mathrm{E}}(O)$ and $q_{\mathrm{E}}(E)$ on the E-page. With it, we get $k_{\mathrm{DE}} = |\vec{r}_{\mathrm{E}}(E)| / |\vec{r}_{\mathrm{D}}(E)|$ since the distance $|\vec{r}_{\mathrm{D}}(E)|$ between the emitter and the eastward detector on the table is known.
- The $k_{\mathrm{ND}}$-frame is adapted to the motion of the N-page: $\vec{u}_{\mathrm{ND}}$ is northward and $k_{\mathrm{DN}} = \vec{r}_{\mathrm{N}}(N) / \vec{r}_{\mathrm{D}}(N)$ obtains from similar measurements on the N-page.

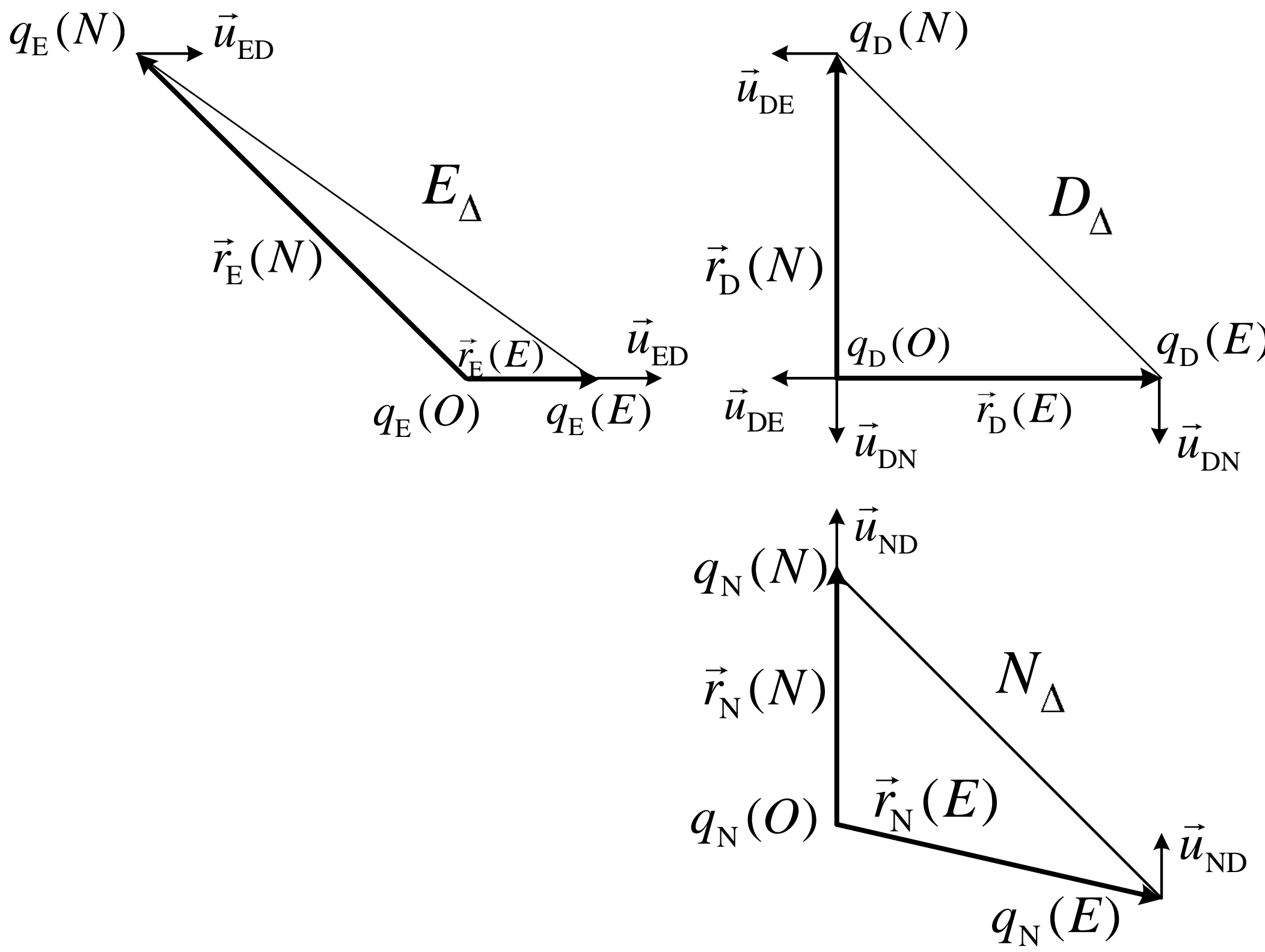

Figure 4. The relations between the triangles prior to the emission event $O$

Very likely, the inertial frames where the E- and the N-page is respectively at rest, the EN-pair of frames, also form a binate frame, but is it really so? A-priori we have no answer: the set-up was not designed to ensure that the relative motion between these two pages is uniform and in a fixed direction. Classical kinematics concludes that the NE-pair does indeed form a binate frame. If we accept this conclusion, it is legitimate to substitute in (1) the classically calculated distances. But the substitution yields two incompatible equations: The classical approach fails! The binate-frame analysis below delivers the correct answer.

All the considered events (the emission and the two detections) are evidently three-place coincidences. For example, to show the eastward detection $E$ in Fig. 5, just slide $E_{\Delta}$ eastward and $N_{\Delta}$ northward until $q_N(E)$ and $q_E(E)$ coincide at $q_D(E)$. Applied at this event, the light propagation relation (1) for the $k_{\mathrm{ND}}$ -frame is:

$$(1-k_{\mathrm{ND}})\left(\left|\vec{r}_{\mathrm{N}}(E)\right|+\left|\vec{r}_{\mathrm{D}}(E)\right|\right)+(1+k_{\mathrm{ND}})\left[\vec{r}_{\mathrm{N}}(E)\bullet\vec{u}_{\mathrm{ND}}+\vec{r}_{\mathrm{D}}(E)\bullet\vec{u}_{\mathrm{DN}}\right]=0\,.$$

Similar equations apply for the other detection events. These are true scalar equation, and thus the choice of coordinate axes is immaterial. Yet, it is convenient to have for the $\mathbb{R}^2$ vectors a basis that is adapted to the diagram: $\vec{1}_x$ eastward and $\vec{1}_y$ northward. With this basis, we get from the above the $\vec{1}_y$-component of $\vec{r}_{\mathrm{N}}(E)$ and from a similar equation the $\vec{1}_x$-component of $\vec{r}_{\mathrm{E}}(N)$. (In the example $\vec{r}_{\mathrm{N}}(E)\bullet\vec{1}_y = -.225$ and $\vec{r}_{\mathrm{E}}(N)\bullet\vec{1}_x = -1.05$, in units of the distance from the emitter to the detector.) Now all the radius vectors to the detection events are fully known since four other radius vectors were measured earlier, to obtain $k_{\mathrm{DE}}$ and $k_{\mathrm{DN}}$.

Consider next the presumed binate frame formed with the EN-pair, for which we know neither the separation rate $k_{\mathrm{EN}}$, nor its unit vectors $\vec{u}_{\mathrm{EN}}$ and $\vec{u}_{\mathrm{NE}}$. What we know are the radius vectors to the eastern and northern detection on the E-page and N-page. To apply (1) for the presumed $k_{\mathrm{EN}}$-frame, we use the copies of $E_\Delta$ and $N_\Delta$ that are already on the desk, which is our choice of a common $\mathbb{R}^2$. Recall that $E_\Delta$ and $N_\Delta$ in the diagram of Fig. 4 are aligned along the unit vectors of the desktop, not those of the E-page and the N-page. (If these unit vectors were indeed correctly aligned, classical kinematics would have obtained the right answer.)

Say we rotate $E_\Delta$ by some angle $\alpha$ to correctly align it with $N_\Delta$. The vectors on the common $\mathbb{R}^2$ (the desktop) would then meet: $\vec{u}_{\mathrm{EN}} = \left.\left(\dfrac{R(\alpha)\vec{r}_{\mathrm{E}} - \vec{r}_{\mathrm{N}}}{\left|R(\alpha)\vec{r}_{\mathrm{E}} - \vec{r}_{\mathrm{N}}\right|}\right)\right|_E = \left.\left(\dfrac{R(\alpha)\vec{r}_{\mathrm{E}} - \vec{r}_{\mathrm{N}}}{\left|R(\alpha)\vec{r}_{\mathrm{E}} - \vec{r}_{\mathrm{N}}\right|}\right)\right|_N$.
These equations yield first, the angle $\alpha$ for the rotation matrix $R(\alpha)$, and then $\vec{u}_{\mathrm{EN}}$. With these, $k_{\mathrm{EN}}$ obtains from (1). (Calculations for the numeric example yield: $\alpha = 5.453^o$, $\vec{u}_{\mathrm{EN}}{}^T = \begin{bmatrix}-.955 & .297\end{bmatrix}$ and $k_{\mathrm{EN}} = .387$.)

The reader might have already noticed that the rotation angle between the E-page and the N-page is the famous Wigner angle. These pages move on the table in northward and eastward directions, which are at a right angle. With $\theta = 90^o$, the Wigner angle $\delta$ obtains from $\cos(\delta) = \dfrac{K^2 - 1}{K^2 + 1}$ (Eq. 26 of reference [9]), where $K = \sqrt{\dfrac{\gamma_{\mathrm{ED}} + 1}{\gamma_{\mathrm{ED}} - 1}}\sqrt{\dfrac{\gamma_{\mathrm{ND}} + 1}{\gamma_{\mathrm{ND}} - 1}}$ (Eq. 7 of the same reference). Each $\gamma$-factor is calculated from the respective separation rate, $k_{\mathrm{ED}}$ and $k_{\mathrm{ND}}$, since as earlier mentioned: $\gamma = \dfrac{1 + k^2}{2k}$. (For our numeric example, we got the expected $\delta = 5.453^o$, that is, the earlier calculated value of $\alpha$.)

## VIII. HAMILTONIAN DYNAMICS OF MULTI-PARTICLE SYSTEMS

Classical Hamiltonian dynamics views a closed system of $N$ -particles as evolving along a path in phase space. In this geometric space, four position-coordinates $q_i^{\ \mu}$ and their conjugate momentum-coordinates $p_{i\mu}$ represent the $i$ -particle. The motion of this particle is determined by canonical equations: $\dfrac{dq_i^{\ \mu}}{d\xi} = \dfrac{\partial H}{\partial p_{i\mu}}$ and $\dfrac{dp_{i\mu}}{d\xi} = -\dfrac{\partial H}{\partial q_i^{\ \mu}}$, derived from the system's Hamiltonian, $H(q,p)$. Here, $q$ stands for all the $q_i^{\ \mu}(\xi)$, and $p$ for all the $p_{i\mu}(\xi)$. The particle's path parameter $\xi$ is associated with the specified Hamiltonian.

We often use the canonical coordinates of a particle as its position-coordinates in phase space: $q_i^{\ 1} = x_{\mathrm{A}i}$, $q_i^{\ 2} = x_{\mathrm{B}i}$, $q_i^{\ 3} = y_{\mathrm{A}i} = -y_{\mathrm{B}i}$, $q_i^{\ 4} = z_{\mathrm{A}i} = z_{\mathrm{B}i}$, Then to be physically meaningful, a Hamiltonian should retain its functional form and value when changing the binate frame and/or changing its canonical parameterization. This ensures that also the path parameter $\xi$ keeps its values invariant under such changes.

### A. Dynamics of a free particle

In Sec. VI, we called a particle free when the separation between near events along its path through phase space was maximal. Dynamically, a particle is *free* particle when its motion is not affected by any external factors. Then its Hamiltonian depends only on an invariant scalar measure of its momentum. The simplest momentum invariant has the form: $\|\mathrm{p}\|^2 = \eta^{\mu\nu} p_{i\mu} p_{i\nu}$, where $\eta^{\mu\nu}$ is the inverse of the earlier defined $\eta_{\mu\nu}$. In any $k_{\mathrm{AB}}$ -frame, the non-null entries of $\eta^{\mu\nu}$ are: $\eta^{12} = \eta^{21} = \tfrac{1}{2}\left(k_{\mathrm{AB}}^{\ \ -1} + k_{\mathrm{AB}}\right)$ and $\eta^{\mu\nu} = -1$ for $\mu = \nu$. It is convenient to use the short hand: $q_\mu = \eta_{\mu\nu} q^\nu$ and $p^\mu = \eta^{\mu\nu} p_\mu$.

A free particle of mass $m$ is a system of $m$ co-moving particles of unit mass, and its momentum is an additive property (component by component). Thus the momentum of a particle-cum-system has a magnitude proportional to the particle's mass: $\|\mathrm{p}\| = mc$, where $c$ is here a constant to convert between mass and momentum units. (The speed of light $c$ fulfils the same purpose in special relativity; hence, we use the same letter.)

The motion of a free particle obtains with the Hamiltonian: $H = \dfrac{1}{2}\dfrac{\|\mathrm{p}\|^2}{mc}$. The factor $\tfrac{1}{2}$ is for convenience, but presenting the Hamiltonian in units of momentum, rather than of energy, ensures that the evolution parameter $\xi$ emerges in units of length.. This seems appropriate here since neither a notion of 'time' nor one of 'energy' have been introduced yet for the binate frame approach. The canonical equations: $\dfrac{dq^\mu}{d\xi} = \dfrac{1}{mc} p^\mu$ and $\dfrac{dp_\mu}{d\xi} = 0$, are met with: $q^{\mu} = \frac{1}{mc} p^{\mu}\big|_{\xi=0} \xi + q^{\mu}\big|_{\xi=0}$. If the free particle is at rest on the A-frame of an adapted $k_{\mathrm{AB}}$-frame: $p^1 = 0$ and $p^2 = -\frac{1-k_{\mathrm{AB}}{}^2}{2k_{\mathrm{AB}}} mc$.

Consider a free particle moving along an AB-line in some $k_{\mathrm{AB}}$-frame. If its motion is symmetric relative to the paired inertial frames of the $k_{\mathrm{AB}}$-frame: $p^1 = p^2 = -\frac{1-k_{\mathrm{AB}}}{2\sqrt{k_{\mathrm{AB}}}} mc$. Otherwise, the particle's inertial frame, the P-frame, forms an adapted $k_{\mathrm{PA}}$-frame with the A-frame, and yet another adapted $k_{\mathrm{PB}}$-frame with the B- frame. The momentum of this free particle is: $p^1 = \frac{1-k_{\mathrm{PA}}{}^2}{2k_{\mathrm{PA}}} mc$ and $p^2 = -\frac{1-k_{\mathrm{PB}}{}^2}{2k_{\mathrm{PB}}} mc$.

The above relations still hold true even for a particle that moves on an AB-line under the influence of external factors, though then $k_{\mathrm{PA}}$ and $k_{\mathrm{PB}}$ are not constant anymore since the particle's motion is not uniform. Notwithstanding, at all the events attended by the particle: $Min(k_{\mathrm{AB}}, k_{\mathrm{PA}}, k_{\mathrm{PB}}) = \sqrt{k_{\mathrm{AB}} k_{\mathrm{PA}} k_{\mathrm{PB}}}$, that is, the smallest $k$-value of the three is the product of the other two. Replacing the $k$'s in terms of the $\beta$'s yields the usual special relativistic addition of collinear velocities.

## B. Dynamics of a particle on a circular track

Consider a particle moving on a circular track of radius $R$, but otherwise free from factors that affect its motion. We then expect the particle to meet canonical equations derived from: $H = \frac{1}{2}\frac{p_{\mu} p^{\mu}}{mc} - \frac{1}{2}\lambda\left(M_{\mu\nu} q^{\mu} q^{\nu} - R^2\right)$, where the first term comes from the free particle's Hamiltonian and the second term, from the track's constraint. For the latter, we used a Lagrangian multiplier $\lambda$ and a symmetric matrix $M_{\mu\nu}$.

In a $k_{\mathrm{AB}}$-frame adapted to the particle's motion: The A-frame is the inertial frame where the track is at rest in the $x_{\mathrm{A}} y_{\mathrm{A}}$-plane, with the centre at the origin of this frame's canonical coordinates. The B-frame is the inertial frame where the particle is at rest just prior to entering the circular track. The track imposes the constraint: $x_{\mathrm{A}}{}^2 + y_{\mathrm{A}}{}^2 = R^2$ on the particle's coordinates. Thus in the adapted $k_{\mathrm{AB}}$-frame, the only non-null entries of $M_{\mu\nu}$ are: $M_{11} = M_{33} = 1$.

The particle enters the track at the event with the coordinates: $x_{\mathrm{A}} = x_{\mathrm{B}} = 0$ and $y_{\mathrm{A}} = -R$. (The $z$-coordinate is irrelevant here and was omitted.) Its initial momentum has just one non-null component: $p^1 = -\frac{1-k_{\mathrm{AB}}{}^2}{2k_{\mathrm{AB}}} mc$.

The canonical equations have a solution for $\omega R = \frac{1-{k_{AB}}^2}{2k_{AB}}$, where $\omega$ is the angular rate of change (per unit length): $x_A = -R\sin(\omega\xi)$, $x_B = -\frac{1+{k_{AB}}^2}{2k_{AB}} R\left[\omega\xi - \sin(\omega\xi)\right]$ and $y_A = -R\cos(\omega\xi)$. Simple calculations show that $s = \xi$ at all the events on the particle's path through phase space. The Hamiltonian's constant value is $\frac{1}{2}mc$.

If a sensor on the track marks the B-frame each time it detects the particle, any two consecutive marks would be separated by a distance: $\frac{1+{k_{AB}}^2}{2k_{AB}} 2\pi R$, which is greater than the length of the circular track.

With the appropriate substitutions, inertial observers could obtain from (4) the times of the events attended by the particle: An observer on the A-frame gets $ct_A = \gamma_{AB} s$. An observer on the B-frame gets $ct_B = \gamma_{AB}(\beta_{AB} x_A + ct_A)$, which is just as expected from the Lorentz transformation between the two inertial observers.

### C. Mass-ratio derived from inelastic collisions

We explained in Sec. VI that after a fully inelastic collision of identical particles, the coalesced particle moves symmetric relative to the inertial frames where the incoming particles were initially at rest. When the incoming particles do not have the same mass, the coalesced particle that emerges from the collision moves asymmetric. This raises the possibility of measuring kinematically the mass-ratio of two particles, as illustrated next.

We consider here two free particles in relative motion. In a $k_{AB}$-frame adapted to this motion, say $M$ is the mass of the particle initially at rest on the A-frame and $m$ is the mass of the particle initially at rest on the B-frame. Before the collision, the particles are free and thus: ${p_M}^1 = 0$, ${p_M}^2 = -\frac{1-{k_{AB}}^2}{2k_{AB}} Mc$, ${p_m}^1 = -\frac{1-{k_{AB}}^2}{2k_{AB}} mc$, and ${p_m}^2 = 0$. After the collision, the two particles coalesce and move together. A balance of momentum yields: $p^1 = -\frac{1-{k_{AB}}^2}{2k_{AB}} mc$ and $p^2 = -\frac{1-{k_{AB}}^2}{2k_{AB}} Mc$, and thus $\|\mathrm{p}\|^2 = (mc)^2 + \frac{1+{k_{AB}}^2}{k_{AB}} mMc^2 + (Mc)^2$ for the coalesced particle. Since the particle's momentum is proportional to its mass, the above relation shows that $\|\mathrm{p}\| = \alpha(m+M)$, where $\alpha = \sqrt{1 + \frac{(1-k_{AB})^2}{2k_{AB}} \frac{mM}{(m+M)^2}}$. Clearly, the mass of the composite particle is larger than the sum of the two masses.

Evidently, precautions have to be taken to ensure that the collision is fully inelastic yet conserves the energy of the two-particle system. Say a (massless) spring is inserted between the two particles. The compression of the spring cushions the impact and thus captures the energy that otherwise would have dissipated in the environment. The mass equivalent of this energy is: $(\alpha-1)(m+M)$.

As an alternative to the above calculations for the coalesced particle's momentum, recall from our earlier kinematical analysis that: $\frac{dx_{\mathrm{A}}}{ds}=-\frac{1-{k_{\mathrm{PA}}}^2}{2k_{\mathrm{PA}}}$ and $\frac{dx_{\mathrm{B}}}{ds}=-\frac{1-{k_{\mathrm{PB}}}^2}{2k_{\mathrm{PB}}}$. We substitute them in the canonical equations to obtain: $p^1=-\frac{1-{k_{\mathrm{PA}}}^2}{2k_{\mathrm{PA}}}\alpha(M+m)c$ and $p^2=-\frac{1-{k_{\mathrm{PB}}}^2}{2k_{\mathrm{PB}}}\alpha(M+m)c$. These are matched to the here obtained components of the coalesced particle: $\frac{1-{k_{\mathrm{AB}}}^2}{2k_{\mathrm{AB}}}m=\frac{1-{k_{\mathrm{PA}}}^2}{2k_{\mathrm{PA}}}\alpha(M+m)$ and $\frac{1-{k_{\mathrm{AB}}}^2}{2k_{\mathrm{AB}}}M=\frac{1-{k_{\mathrm{PB}}}^2}{2k_{\mathrm{PB}}}\alpha(M+m)$. Divide the two relations, side by side, to get the mass-ratio $\frac{m}{M}=\frac{k_{\mathrm{PB}}\left(1-{k_{\mathrm{PA}}}^2\right)}{k_{\mathrm{PA}}\left(1-{k_{\mathrm{PB}}}^2\right)}$ in terms of measurable $k$-values.

## D. Bertozzi experiments with relativistic electrons

The relativistic relation between the speed of a particle and its kinetic energy differs from that obtained by classical mechanics. The first evidence in support of the relativistic relation came from Bertozzi experiments with high-energy electrons [10]. A binate frame analysis of the relevant aspects of the reported experiments reaches the same conclusion.

In an adapted $k_{\mathrm{AB}}$-frame, a calorimeter of mass $M$ on the A-body and a particle of mass $m$ on the B-body move toward each other until the calorimeter eventually capture the particle. Thereafter, they move together, with $\|\mathrm{p}\|^2=(mc)^2+\frac{1+{k_{\mathrm{AB}}}^2}{k_{\mathrm{AB}}}mMc^2+(Mc)^2$, as earlier calculated. This yields $\frac{1}{2}\frac{\|\mathrm{p}\|^2}{(m+M)}=\frac{1}{2}(m+M)c^2+Q_k$, where $Q_k$ accounts for the energy equivalent of the heat generated by the mechanical process that stopped the particle: $Q_k=\frac{(1-k_{\mathrm{AB}})^2}{2k_{\mathrm{AB}}}\frac{mMc^2}{m+M}\simeq\frac{1}{2}\left(\sqrt{{k_{\mathrm{AB}}}^{-1}}-\sqrt{k_{\mathrm{AB}}}\right)^2mc^2$, to the first order in $\frac{m}{M}\ll 1$.

Now recall that $\beta_{\mathrm{AB}}=\frac{1-{k_{\mathrm{AB}}}^2}{1+{k_{\mathrm{AB}}}^2}$ is the special relativistic (fractional) speed of the particle in the frame of laboratory with the calorimeter. Since both $Q_k/mc^2$ and $\beta_{\mathrm{AB}}$ depend on the same parameter $k_{\mathrm{AB}}$, we could plot them one against the other. With our results, the plot traces Bertozzi's experimental curve that relates the particle speed, $\beta_{\mathrm{AB}}$, to its kinetic energy, $Q_K/mc^2$, measured as the energy captured in the calorimeter. The two curves are not identical by accident! In the expression for $Q_k$, substitute $k_{\mathrm{AB}}$ in terms of the Lorentz factor, $\gamma_{\text{AB}} = \dfrac{1+k_{\text{AB}}{}^2}{2k_{\text{AB}}}$. This yields $Q_K/mc^2 = (\gamma_{\text{AB}} - 1)$, the relativistic kinetic energy per unit of the rest mass energy of a particle moving with $\beta_{\text{AB}}$ relative to a laboratory.

## IX. DYNAMICS OF A MULTI-BODY SYSTEM

We consider the motions involved in experiments where two bodies, each identified by its mass, $m$ and $\overline{m}$, interact by mechanical means as they slide on frictionless tracks mounted on a much more massive body. The two bodies are connected as in Fig. 5 by a (weightless) rigid rod of length $L$. A cart (represented by $O_{\text{B}}$) carries $\overline{m}$ until it enters the horizontal track from the right. At this initial event, $\overline{m}$ catches the rod's lower end and starts pulling downward $m$ which is attached to the rod's upper end. This slows the motion of $\overline{m}$; it begins to fall behind the cart, whose uniform motion is unchanged.

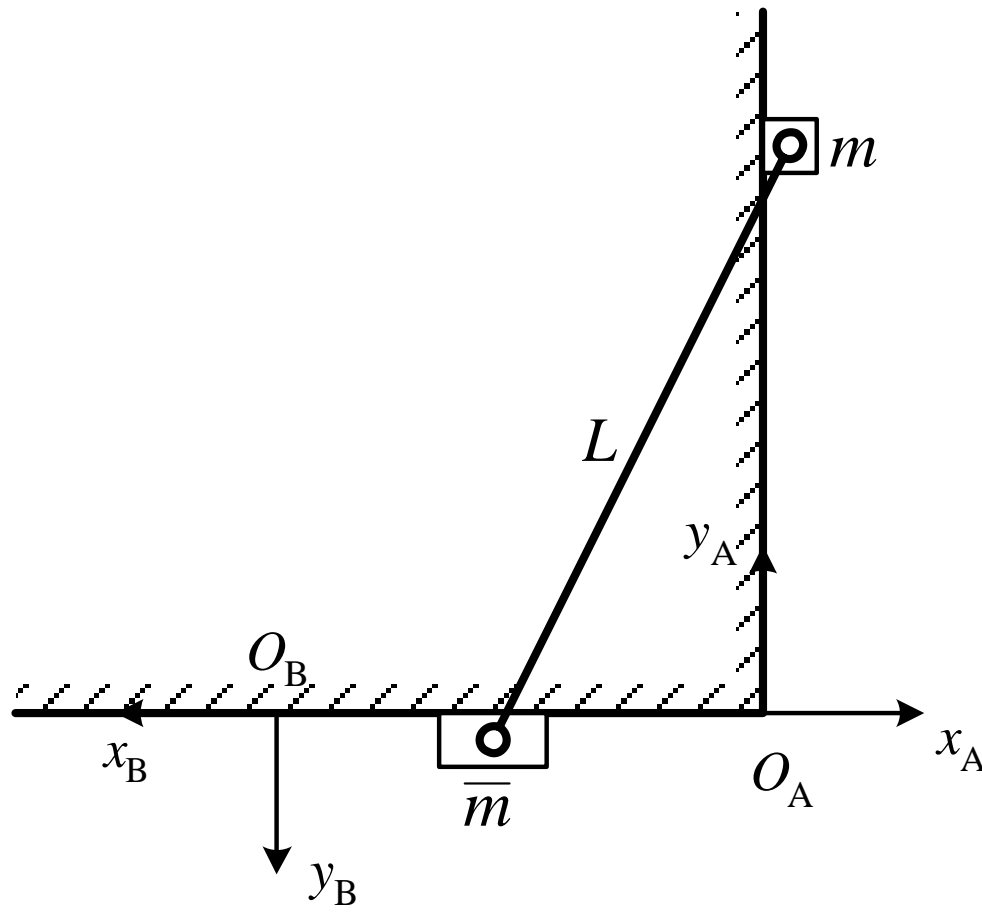

Figure 5. Two bodies sliding on tracks and connected by a rigid rod

This classical two-body system is interesting because the planar motion of the rod is typical of how a rigid body moves. Predictions are, in principle, easily verified with an emitter somewhere on the horizontal track, in the path of $\overline{m}$, and a detector on $m$ itself. When $\overline{m}$ reaches the emitter, say it triggers the emission of a light that is detected by $m$. If the detection event leaves a mark on the vertical track, then the distance from the mark to the horizontal track should match its predicted value.

Our analysis uses an adapted binate frame: The A-frame is the body with the tracks; $m$ is initially at rest on this inertial frame and $O_{\text{A}}$ is where the tracks meet. The B-frame is the cart; $\overline{m}$ is initially at rest on this inertial frame, at $O_{\text{B}}$. The tracks impose: $x_{\text{A}} = 0$ and $\overline{y}_{\text{B}} = -\overline{y}_{\text{A}} = 0$ on the considered motion. This starts at $\xi = 0$, when $\overline{m}$ catches the rod's lower end. Initially, $m$ is at $x_{\text{B}} = 0$ and $y_{\text{A}} = -y_{\text{B}} = L$, with $dy_{\text{A}}/d\xi = 0$, whereas $\overline{m}$ is at $\overline{x}_{\text{A}} = \overline{x}_{\text{B}} = 0$, with $d\overline{x}_{\text{B}}/d\xi = 0$.

It is important to highlight the simplifications brought about by the usual assumptions made in classical mechanics when analyzing situations similar to that described here:

- The tracks are frictionless, and so the system is conservative;
- The cart remains an inertial frame no matter what happens to $\overline{m}$;
- The rod is weightless and thus contributes nothing to the energy of the system;
- The path parameter $\xi$ provides a dynamic notion of simultaneity;
- The rod is rigid and, at $\xi = 0$, it starts to affect the two bodies simultaneously.

The position-coordinates are: $q^1 = x_{\mathrm{A}}$, $q^2 = x_{\mathrm{B}}$, $q^3 = y_{\mathrm{A}} = -y_{\mathrm{B}}$, $\overline{q}^1 = \overline{x}_{\mathrm{A}}$, $\overline{q}^2 = \overline{x}_{\mathrm{B}}$, and $\overline{q}^3 = \overline{y}_{\mathrm{A}} = -\overline{y}_{\mathrm{B}}$, with $p_\mu$ and $\overline{p}_\mu$, for $\mu = 1,2,3$, the respective conjugate momenta. The selected Hamiltonian has a form similar to those encountered earlier:

$$H = \frac{p_\mu p^\mu}{2mc} + \frac{\overline{p}_\mu \overline{p}^\mu}{2\overline{m}c} - \frac{\lambda}{2}\left[ L^2 - M_{\mu\nu}\left(\overline{q}^\mu - q^\mu\right)\left(\overline{q}^\nu - q^\nu\right)\right].$$

In the specified adapted $k_{\mathrm{AB}}$-frame, the non-null entries of the matrix $M_{\mu\nu}$ that enforces the rod rigidity are: $M_{11} = M_{33} = 1$.

The canonical equations admit a solution with angular frequencies $\omega = \sqrt{\dfrac{\lambda}{mc}}$ and $\overline{\omega} = \sqrt{\dfrac{\lambda}{\overline{m}c}}$ (per unit length), where $\lambda$ obtains from: $\dfrac{1-k_{\mathrm{AB}}{}^2}{2k_{\mathrm{AB}}}\sin\left(\overline{\omega}\xi\right) = \overline{\omega}L\sin\left(\omega\xi\right)$. The rod's rotation is non-uniform since $\lambda$ depends on $\xi$. The solution is valid for $\xi \geq 0$ until it reached $\xi_f = \dfrac{\pi}{2\omega}$, when $m$ arrives at the bottom of its track.

$$x_{\mathrm{A}} = 0, \quad x_{\mathrm{B}} = -\frac{1-k_{\mathrm{AB}}{}^2}{2k_{\mathrm{AB}}}\xi, \quad y_{\mathrm{A}} = -y_{\mathrm{B}} = L\cos\left(\omega\xi\right),$$

$$\overline{x}_{\mathrm{A}} = -L\sin\left(\omega\xi\right), \quad \overline{x}_{\mathrm{B}} = \frac{1+k_{\mathrm{AB}}{}^2}{2k_{\mathrm{AB}}}\left[x_{\mathrm{B}} + L\sin\left(\omega\xi\right)\right], \quad \overline{y}_{\mathrm{A}} = -\overline{y}_{\mathrm{B}} = 0.$$

A given $\xi$ selects two distant events attended by $\overline{m}$ and $m$, respectively, that here are viewed as being *simultaneous* for the system. Thus for the considered system, $\xi = ct$ is an invariant Newtonian time (in units of length). When substituted in our solution, it yields: $x_{\mathrm{A}} = 0$, $y_{\mathrm{A}} = L\cos\left(\omega_c t\right)$, $\overline{x}_{\mathrm{A}} = -L\sin\left(\omega_c t\right)$ and $\overline{y}_{\mathrm{A}} = 0$, where $\omega_c = \sqrt{\lambda_c / m}$ and $\overline{\omega}_c = \sqrt{\lambda_c / \overline{m}}$ (now reckoned per unit time) meet $\dfrac{1-k_{\mathrm{AB}}{}^2}{2k_{\mathrm{AB}}}\sin\left(\overline{\omega}_c t\right) = \dfrac{\overline{\omega}_c L}{c}\sin\left(\omega_c t\right)$. It is easily verified that this is also the classical mechanics solution in the inertial A-frame.

However, our solution also yields: $x_{\mathrm{B}} = -\dfrac{1-k_{\mathrm{AB}}{}^{2}}{2k_{\mathrm{AB}}}ct$ and $\overline{x}_{\mathrm{B}} = \dfrac{1+k_{\mathrm{AB}}{}^{2}}{2k_{\mathrm{AB}}}\left[x_{\mathrm{B}} + \overline{x}_{\mathrm{A}}\right]$. This is unlike classical mechanics: $\overline{x}_{\mathrm{B}}$ is greater than $x_{\mathrm{B}} + x_{\mathrm{A}}$ by a Lorentz factor, $\gamma_{\mathrm{AB}} = \dfrac{1+k_{\mathrm{AB}}{}^{2}}{2k_{\mathrm{AB}}}$, as expected from special relativity.

The appropriate substitutions in (4) yield the following special relativistic times: $ct_{\mathrm{A}} = \xi$, $ct_{\mathrm{B}} = \gamma_{\mathrm{AB}}\xi$, $c\overline{t}_{\mathrm{A}} = \gamma_{\mathrm{AB}}\xi$, and $c\overline{t}_{\mathrm{B}} = \gamma_{\mathrm{AB}}\left(ct_{\mathrm{B}} + \beta_{\mathrm{AB}}x_{\mathrm{B}}\right)$. Not unexpectedly, these meet Lorentz transformations. The dynamical notion of system-dependent simultaneity of distance events is evidently not carried over in the spacetime coordinates of either inertial observer: substitutions show that only the initial events (at $\xi = 0$) for the two bodies are viewed as being simultaneous in special relativity.

## X. MASS-AND-SPRING SYSTEM

A new approach is expected to occasionally bring new solutions to old problems. This is illustrated below with an analysis of a familiar mass-and-spring system. What makes it interesting is the fundamental role it plays in modern physics: its relativistic quantum mechanics is still being explored in peer-reviewed articles, [11, 12] more than half a century after relativistic analyses of the system's dynamics were first published.

Classical mechanics shows that a mass $m$ attached to a linear spring of nominal null mass and constant $\kappa_s$ oscillates back and forth harmonically, with an angular frequency $\omega = \sqrt{\kappa_s / m}$, when released from a position where the spring is not fully relaxed. Special relativistic calculations predict anharmonic oscillations, but not all published solutions yield the same oscillations. [5] Our binate frame analysis next predicts the particle's path in the phase space of the system. For comparison with special relativity, we also present our solution in terms of spacetime coordinates. It shows that in the inertial frame where the mass-and-spring system is at rest (as a whole), the motion is harmonic with energy-dependent angular frequency. The motion is anharmonic in any other inertial frame.

### A. Binate frame analysis

The mass-and-spring system that we consider here is mounted on a moving cart. One end of the spring is attached to the cart; the mass is at the other end, initially at a distance $L$ from where the spring is fully relaxed. A uniform motion is imparted to the cart so that the mass, when free to oscillate back and forth, is once in each period momentarily at rest relative to the ground, and marks its position there. The distance on the ground between any two consecutive marks is then the wavelength of the classically predicted harmonic oscillations: $\lambda_C = 2\pi L$. This prediction is accurate for slow oscillations: $\beta = \omega L / c \ll 1$. In contrast, our dynamical analysis below encompasses the full range of motions.

We first adapt a $k_{\mathrm{AB}}$-frame to the mass-and-spring system: the A-frame is the cart and the B-frame is the ground. The canonical coordinates on the cart are located at the place where the spring is fully relaxed. On the ground, the coordinates are located where the mass placed its first mark, at the event $O(x_{\mathrm{A}}=0, x_{\mathrm{B}}=0)$. We use $(q^1 = x_{\mathrm{A}}, q^2 = x_{\mathrm{B}})$ for the position-coordinates and $(p_1, p_2)$ for their conjugate momentum-coordinates.

The state of a deformed spring depends on how far along the $x_{\mathrm{A}}$-coordinate the mass is from the place at $x_{\mathrm{A}}=0$ where the spring is fully relaxed. In each period, the spring is fully relaxed at two events and at one of them, the mass rests momentarily relative to the ground: $dq^2/d\xi = 0$. If then the mass with the relaxed spring attached to it were to be released from the connection with the cart, the two would remain at rest on the ground. The energy of the so released mass-and-relaxed-spring would be just the energy of the free particle since a relaxed spring holds no energy.

With the mass at $x_{\mathrm{A}}$, the energy of the linear spring is $\frac{1}{2}\kappa_s x_{\mathrm{A}}^2$. The spring is either fully compressed or fully extended when the mass is at $q^1 = \pm L$ and $dq^1/d\xi = 0$. At these events, the energy of the system is larger that of a free mass momentarily at rest on the cart by an amount equal to the energy of the fully deformed spring, $\frac{1}{2}\kappa_s L^2$.

The considerations of the last two paragraphs suggest we use a Hamiltonian of the form $H = \dfrac{1}{2}\left(\dfrac{p_\mu p^\mu}{m} - K_{\mu\nu}q^\mu q^\nu\right)$, in units of energy, whose constant value is the energy of the initially free particle: $\dfrac{1}{2}mc^2$. In the adapted $k_{\mathrm{AB}}$-frame, the spring's matrix $K_{\mu\nu}$ has just one non-null entry: $K_{11} = \kappa_s$. The canonical equations of motion are met by:

$$\begin{cases} x_{\mathrm{A}} = q^1 = -L\sin(\omega\xi) \\ p_1 = -\left[\left(\dfrac{1+k_{\mathrm{AB}}{}^2}{1-k_{\mathrm{AB}}{}^2}\right)^2 - \cos(\omega\xi)\right] m\omega L \end{cases} \quad \text{and} \quad \begin{cases} x_{\mathrm{B}} = q^2 = -\dfrac{1+k_{\mathrm{AB}}{}^2}{2k_{\mathrm{AB}}} L\left[\omega\xi - \sin(\omega\xi)\right] \\ p_2 = -\dfrac{2k_{\mathrm{AB}}}{1+k_{\mathrm{AB}}{}^2}\left(\dfrac{1+k_{\mathrm{AB}}{}^2}{1-k_{\mathrm{AB}}{}^2}\right)^2 m\omega L \end{cases}.$$

An examination of the solution shows that the particle at $\xi = 0$ is indeed at rest on ground, with the spring fully relaxed. But a simple calculation shows that only when the motion of the cart, which is specified with $k_{\mathrm{AB}}$, and the spring's maximum extension $L$ meet $k_{\mathrm{AB}}{}^{-1} - k_{\mathrm{AB}} = 2\omega L/c$ is the momentum of the particle at $\xi = 0$ that of a particle at rest momentarily on the ground: $p^1 = -\dfrac{1-k_{\mathrm{AB}}{}^2}{2k_{\mathrm{AB}}}mc$ and $p^2 = 0$. In principle at least, this condition is verifiable during the set-up of the experiment.

The marks left on the ground are essential records of the experimental outcome. The first mark is put at $\xi = 0$, and the next mark, when $\omega\xi = 2\pi$. The distance between them is: $\lambda_R = \pi\left(k_{\mathrm{AB}}{}^{-1} + k_{\mathrm{AB}}\right)L$. This is the wavelength of the ensuing oscillations: subsequent marks on the ground result from events separated by the same $\Delta\xi = 2\pi/\omega$ interval. Thus $\lambda_R = \gamma_{\mathrm{AB}}\lambda_C$, that is the classical and relativistic predictions differ by the Lorentz factor.

## B. The binate frame solution converted to spacetime coordinates

The binate-frame solution could be presented in spacetime coordinates. The special relativistic time coordinates on the paired inertial frames of the adapted $k_{\mathrm{AB}}$-frame are: $ct_{\mathrm{A}} = \dfrac{1+k_{\mathrm{AB}}{}^2}{1-k_{\mathrm{AB}}{}^2}\omega\xi L$ and $ct_{\mathrm{B}} = \dfrac{1+k_{\mathrm{AB}}{}^2}{2k_{\mathrm{AB}}}\left(\dfrac{1+k_{\mathrm{AB}}{}^2}{1-k_{\mathrm{AB}}{}^2}\omega\xi - \dfrac{1-k_{\mathrm{AB}}{}^2}{1+k_{\mathrm{AB}}{}^2}\sin(\omega\xi)\right)L$, from (4).

For an observer on the cart, where the mass-and-spring system as a whole is at rest, the evolution parameter $\xi$ and the calculated coordinate-time $t_{\mathrm{A}}$ are linearly related. It is easily verifiable that the classical Newtonian equation: $m\dfrac{d^2x_{\mathrm{A}}}{dt_{\mathrm{A}}{}^2} = -\omega_{\mathrm{A}}{}^2 x_{\mathrm{A}}$ is met in the frame of this observer. Thus the mass moves harmonically, but $\omega_{\mathrm{A}} = \omega/\gamma_{\mathrm{AB}}$ depends not only on the early mentioned classical $\omega$, but also on the Lorentz factor $\gamma_{\mathrm{AB}}$ calculated with the mass maximum speed relative to the cart, which is also the uniform speed of the cart relative to the ground.

The situation is more complicated for an observer on the ground since $\xi$ and $t_{\mathrm{B}}$ are not linearly related. In this observer's frame, the mass does not oscillate harmonically.

The asymmetry between the expectations of the two observers is a consequence of the asymmetry of the experiment: as a whole, the mass-and-spring system is at rest only on the cart, not on the ground!

## C. A special relativistic Hamiltonian that leads directly to our spacetime solution

It might interest the reader to have a Hamiltonian that special relativity could use to arrive directly at the spacetime form of our solution. We proceed next to convert the binate frame Hamiltonian to its counterpart in spacetime.

Say the inertial frame where the mass-and-spring system is at rest is viewed as being stationary: $x^0 = ct_{\mathrm{A}}$ and $x^1 = x_{\mathrm{A}}$. We use $q^0 = x^0$ and $q^1 = x^1$ for the position-coordinates in phase space, and $p_\mu = [p_0 \quad p_1]$ for the momentum-coordinates. It seems appropriate to specify a Hamiltonian that reflects the energy balance of the system. For this, the simplest Lorentz invariant derived from the momentum alone, $|\mathrm{p}|^2 = p_\mu p^\mu = p_0{}^2 - p_1{}^2$, is used as a measure of the total energy of the system, $\dfrac{1}{2m}p_\mu p^\mu$. As before, $\dfrac{1}{2}K_{\mu\nu}q^\mu q^\nu$ is used to measure the energy of the spring in its compressed state. In the selected inertial frame, the matrix $K_{\mu\nu}$ has one non-null entry: $K_{11} = \kappa_s$. It is not difficult to verify that the Lorentz scalar invariant: $H = \dfrac{1}{2m} p_\mu p^\mu - \dfrac{1}{2} K_{\mu\nu} q^\mu q^\nu$, yields the spacetime version of the binate frame solution. (The corresponding Lagrangian is the one presented in the introductory section.)

The form of our Hamiltonian differs substantially from: $H = \sqrt{(pc)^2 + (mc)^2} + V(x)$, which is the Hamiltonian most often used to study the evolution path $x(t)$ of a particle subject to a potential $V(x)$. Our Hamiltonian also differs from the one used for the phase space evolution of particle ensembles: $H = \sqrt{(pc)^2 + \left(mc^2 + \frac{1}{2}\kappa_s x^2\right)^2}$ [5].

Given the many distinct Hamiltonians for mass-and-spring system proposed over the years, it seems to us highly desirable to submit the different predictions to high-energy experiments, notwithstanding the inherent difficulties of the required set-up.

## XI. CONCLUSION

It is often the case that our understanding of the physical world is enriched when we examine a situation from different points of view. A most vivid example is the emergence of quantum mechanics, not from Newtonian mechanics but from Hamiltonian dynamics, which itself was derived from the Lagrangian approach. It is also the case that those who challenge some not so self-evident, or perhaps only weakly justified assumptions, do feel rewarded when often from their efforts, new and exciting ideas come to light. The famous Euclid's fifth postulate, the parallel postulate, withstood the challenge of many eminent mathematicians for two thousand years before non-Euclidean geometries, so fundamental to modern physics, were first envisaged. History has probably erased from its old records what actually motivated the search for alternatives to Newtonian mechanics or Euclidian geometry. But somewhere, somebody launched the now long forgotten initial challenge to the deeply entrenched view of the day. And most likely, at the time and in the eyes of those who held precious that prevailing view, the challenge was not and, in fact, could never have been fully justified.

In this article, we presented the fundamentals of a binate-frame approach to classical mechanics that provides a timeless alternative to special relativity. Remarkably, not only photons but also massive particles are never at rest in a binate frame. Also remarkable is the absence of clocks, and the resulting lack of any kinematic notion of simultaneity of distant events. We illustrated the approach with examples of applications to familiar situations that were first explained only during the past century, from special relativistic analyses anchored on Lorentz transformations. We think that the binate-frame kinematics is much simpler that its special relativistic counterpart and furthermore, it alone enables the routine application of Hamiltonian dynamics, since no time-coordinates are needed.

There are other remarkable implications worthy of mention at the conclusion of this paper: It is evident form our presentation that all manifestly covariant aspects of special relativity retain their validity in a binate frame approach, but the same could not be said of the observer-based notions, since these emerge from a space-and-time foliation of the events manifold. For example, in a binate frame approach, as in special relativity:

- Mass (in appropriate physical units) is a scalar invariant of the four-momentum;
- Electrodynamics is based on a four-potential $A_\mu$ and the derived field $F_{\mu\nu}$.

But only in special relativity, inertial observers find it meaningful:

- To form Lorentz four-vectors with the classical three-momentum and the total energy (rest-mass + kinetic);
- To split $A_\mu$ into scalar and vector parts, and $F_{\mu\nu}$ into electric and magnetic parts.

The reader might find it useful to go beyond the classical mechanics subject matter presented in this article and envisage the binate frame approach being pursued say in the quantum mechanics domain. For example, the usual quantization of a free particle yields not the time-dependent Schrödinger equation, but a time-free, four-dimensional, second-order differential equation of the Klein-Gordon type. We are inclined to conclude that the binate-frame approach has remarkable implications even elsewhere in modern physics.